\documentclass[reprint,twocolumn,secnumarabic, nobibnotes,showkeys]{revtex4-2}
\usepackage{bm,chemformula,graphicx,titlesec}
\usepackage[unicode]{hyperref}
\hypersetup{colorlinks= true,
	citecolor= blue,
	linkcolor= red,
	urlcolor=magenta,
	filecolor=cyan,
	linkbordercolor={1 0 0},
	citebordercolor={0 1 0},
	urlbordercolor={0 1 1},
} 
\usepackage[mathlines]{lineno}
\usepackage[section]{placeins}
\usepackage{tikz}
\usepackage{float,wrapfig}
\usepackage{microtype}
\usepackage[tight]{subfigure}
\setcitestyle{super}
\DeclareUnicodeCharacter{2005}{.}
\begin{document}
	\title{Glycolytic Wave Patterns in a Simple Reaction-diffusion System with Inhomogeneous Influx: Dynamic Transitions}
	\author{Premashis Kumar}
	\email{pkmanager@bose.res.in}
	\affiliation{S. N. Bose National Centre For Basic Sciences, Block-JD, Sector-III, Salt Lake, Kolkata 700 106, India}
	\author{Gautam Gangopadhyay}
	\email{gautam@bose.res.in}
	\affiliation{S. N. Bose National Centre For Basic Sciences, Block-JD, Sector-III, Salt Lake, Kolkata 700 106, India}
	\date{\today}
	\begin{abstract}
		An inhomogeneous profile of chemostatted species generates a rich variety of patterns in glycolytic waves depicted in a Selkov reaction-diffusion framework here. A key role played by diffusion amplitude and symmetry in the chemostatted species profile in dictating the fate of local spatial dynamics involving periodic, quasiperiodic, and chaotic patterns and transitions among them are investigated systematically. More importantly, various dynamic transitions, including wave propagation direction changes, are illustrated in interesting situations. Besides numerical results, our analytical formulation of the amplitude equation connecting complex Ginzburg Landau and Lambda-omega representation shed light on the phase dynamics of the system. This complete study of the glycolytic reaction-diffusion wave is in line with previous experimental results in open spatial reactor and will provide a knowledge about the dynamics that shape and control biological information processing and related phenomena.	
	\end{abstract}
	                 
	\maketitle
	\section{\label{intro}Introduction}
	Glycolysis, a pivotal energy-generating process in a living system, comprises a collection of reactions. Diverse theoretical schemes have been proposed to properly capture the oscillatory behavior of glycolysis~\cite{selkov, Goldbeter3255, HYNNE2001121, Madsen, wolf2000effect}. Remarkably, the Goldbeter model~\cite{Goldbeter3255} was designed to investigate the spatial effects on the dynamics of glycolysis by incorporating diffusion into reaction kinetics. This spatially extended model and its modified form~\cite{ZHANG2007112} demonstrated various spatiotemporal behaviors in glycolysis, ranging from traveling waves to spatiotemporal chaos depending on distinct initial conditions or substrate concentration. These spatiotemporal patterns closely resemble the experimentally observed glycolytic waves~\cite{mair2002spatio}. Later, the impact of the feedback regulation of phosphofructokinase on sustained spatiotemporal pattern of the glycolytic oscillation was investigated experimentally in an open spatial reactor~\cite{BAGYAN200567}. Motivated by this experimental setting, Lavrova et al.~\cite{LAVROVA2009127} introduced an inhomogeneous substrate influx in the Selkov model~\cite{selkov} to explain the experimentally observed waves' behaviors. In this context, with several subsequent investigations~\cite{Lavrova2009PhaseInflux, Verveykochaos}, the dynamical study of glycolytic wave propagation with inhomogeneous substrate supply emerges as a viable model system for elucidating the central aspect of the energy metabolism.  
	
	Inspired by the capability of the spatially extended Selkov model with inhomogeneous substrate supply~\cite{LAVROVA2009127, Lavrova2009PhaseInflux} in elaborating a rich variety of experimentally observed phenomena, we have also chosen the two-variable Selkov model extended by diffusion. The simplicity and clarity of such a two-variable reaction-diffusion equation would provide the scope to generalize and compare this investigation findings to any activator-inhibitor type reaction-diffusion system~\cite{Murray2003MathematicalApplications, epstein1998introduction}. More specifically, we here consider the reversible Selkov model~\cite{reversibleselkoov} to avoid the obscurity of the many irreversible steps in the kinetics of the Selkov model. Thus our formulation is associated with a collection of elementary reversible chemical reactions and is suitable for drawing reasonable parallels between chemical and biological oscillator dynamics and their control parameters. Moreover, we intend to formulate our model as an open chemical system such that the time-independent profile of inhomogeneous flux of species can be visualized as the effect of maintenance by some external chemical reservoirs, and corresponding chemical species are identified as the chemostatted species~\cite{Rao2016NonequilibriumThermodynamics}. Previously, a similar formulation was implemented in the glycolytic model with homogeneous chemostatted species to investigate traveling waves around the Benjamin-Feir instability~\cite{pkgg2}. So the study of the glycolytic wave with an inhomogeneous influx of chemostatted species here would generalize this formulation procedure for any open chemical system.                   
	
	Furthermore, the reaction-diffusion model with spatially heterogeneous chemostatted species inflow might be a better analog for the modeling of phenomena in living systems and experimental situations~\cite{ertl1991oscillatory, Reversalhetro, gilbert2010developmental, PAGE200595} than commonly investigated reaction-diffusion systems of the chemical and biological oscillator with homogeneous chemostatted species inflow~\cite{Falasco2018InformationPatterns, thermodynamicschemifcalwaves, pkgg, pkgg2, pkgg3}. For a spatially inhomogeneous inflow of the chemostatted species, we can think of individual points of the spatial domain as a glycolytic subsystem having a specific homogeneous chemostatted concentration. Then these subsystems can have different temporal oscillatory natures while coupled via self-diffusion. Then due to the existence of multiple independent oscillatory pathways, it will be possible to realize complex oscillations like quasiperiodicity and even a transition to chaos reported in oscillatory chemical reaction experiment~\cite{experimentquasiperiodic}. We would also like to connect the standard analytical framework of amplitude equation~\cite{Cross2009PatternSystems} in the reaction-diffusion system to our motivations for numerical investigation scenarios in this report. On this basis, the association of the dynamical profiles of wave propagation for periodic, quasiperiodic, chaotic behavior to the phase reversal dynamics(neglecting diffusion contribution~\cite{Lavrova2009PhaseInflux}) will be tested.

	The layout of the paper is as follows. We have described the reversible Selkov model starting from the usual kinetic Selkov model in section~\ref{sec:1}. In the next section, the reaction-diffusion model of glycolysis is presented. In section~\ref{selampli}, the amplitude equation representation of the reaction-diffusion glycolysis model is provided. This section also discusses the detailed exploration of amplitude and phase dynamics of the system, lambda–omega representation, plane waves stability, and wavenumber selection. The next section provides information about inhomogeneous chemostatted influx and possible numerical investigation. The detailed results and discussion are made in section~\ref{reldiss}, respectively. Lastly, we have concluded the paper in section~\ref{Con}.
	
	\section{\label{sec:1}Model Descriptions: Connecting Selkov Model with Reversible  Selkov Model}
	E. E. Selkov came up with a simple kinetic model of enzyme reaction to encapsulate the dynamical properties of glycolysis~\cite{selkov}. Exploiting this simple model, he demonstrated the generation of self-oscillations in glycolysis over a certain range of system parameters. We can represent the Selkov model in terms of  the following chemical reactions: 
	\begin{equation}
		\begin{aligned}
			\rho&=1:&\ch{S_1 + ES_{2}^2&<=>[\text{k\textsubscript{1}}][\text{k\textsubscript{-1}}] S_1ES_{2}^2} \\
			\rho&=2:&\ch{S_1ES_{2}^2&<=>[\text{k\textsubscript{2}}] ES_{2}^2 + S_2}\\
			\rho&=3:& \ch{2 S_2 + E&<=>[\text{k\textsubscript{3}}][\text{k\textsubscript{-3}}]
				ES_{2}^2}\\ 
		\end{aligned}
		\label{crn}
	\end{equation}
	with $\rho$ being reaction step label and $k_{\rho}$ being rate constants. Here, $S_1(ATP)$ and $S_2(ADP)$ are the substrate and product, respectively. This kinetic model includes the product activation of the enzyme, i. e., initially inactive free enzyme  $E(phosphofructokinase)$ becomes active after forming a complex $ES_{2}^2$.
	
	The concentration dynamics in eq. \eqref{crn} can be described as 
	\begin{equation}
		\begin{aligned}
				\frac{ds_1}{dt}&=z_1-k_1s_1x_1+k_{-1}x_2,\\
				\frac{ds_2}{dt}&=k_2x_2-{k_3}s_2^2e+k_{-3}x_1-k_2s_2,\\
				\frac{dx_1}{dt}&=(k_{-1}+k_2)x_2-k_1s_1x_1+{k_3}s_2^2e-k_{-3}x_1,\\
				\frac{dx_2}{dt}&={k_1}s_1x_1-(k_{-1}+k_2)x_2,\\
				\frac{de}{dt}&=-k_3	s_2^2e+k_{-3}x_1,		
				\label{dynamic}
		\end{aligned}
	\end{equation}
	where $x_1=[ES_{2}^2],x_2=[S_1ES_{2}^2],s_1=[S_1],s_2=[S_2]$ are concentrations of the chemical species and $e$ is the concentration of the free enzyme. Further, $z_1$ is the substrate supply rate, and $k_2s_2$ is the product removal rate.

	After dimensionless analysis of eq. \eqref{dynamic}, we have 
		\begin{subequations}
			\begin{align}
				\frac{d \zeta_1}{ d\theta}&=z-(1+\frac{k_{-1}}{k_2})\frac{\zeta_1x_1}{e_0}+\frac{k_{-1}}{k_2}\frac{x_2}{e_0},\label{zeta1anotherdynamic}\\
				\frac{d \zeta_2}{d \theta}	&=\alpha_2[\frac{x_2}{e_0}-\frac{k_{-3}}{k_2}\frac{e}{e_0}\zeta_2^2+\frac{k_{-3}}{k_2e_0}x_1-X_2\zeta_2], \label{zeta2anotherdynamic} \\
				\epsilon\frac{d x_1}{ d \theta}&=x_2-x_1\zeta_1-\frac{K_3}{K_1+1}[x_1-\zeta_2^2e],\label{x1anotherdynamic}\\
				\epsilon\frac{d x_2}{d \theta}&=\zeta_1x_1-x_2,
				\label{x2anotherdynamic}\\
				\epsilon\frac{d e}{ d \theta}&=\frac{K_3}{K_1+1}[x_1-\zeta_2^2e],
				\label{enzanotherdynamic}
			\end{align}
			\label{degenerate}
		\end{subequations}
		with $\theta=\frac{k_1k_2e_0t}{k_{-1}+k_2}$, $z=\frac{z_1}{k_2e_0}$, $\alpha_2=\frac{k_2+k_{-1}}{k_1}\sqrt{\frac{k_3}{k_{-3}}}$, $X_2=\frac{1}{e_0}\sqrt{\frac{k_{-3}}{k_3}}$, $\epsilon=\frac{e_0k_1k_2}{(k_{-1}+k_2)^2}$, $K_3=\frac{k_{-3}}{k_2}$, $K_1=\frac{k_{-1}}{k_2}$. $\zeta_1=\frac{k_1}{k_{-1}+k_2}s_1$ and $\zeta_2=\sqrt{\frac{k_3}{k_{-3}}}s_2$ are relative concentrations of substrate  and product, respectively and $e_0=x_1+x_2+e$ is the total enzyme concentration. In eq. \eqref{x1anotherdynamic}, \eqref{x2anotherdynamic}, and \eqref{enzanotherdynamic}, derivatives contain a factor $\epsilon$ which can be very small. Under the condition, $\epsilon\ll1$, the system in eq. \eqref{degenerate} can be approximated by eq. \eqref{zeta1anotherdynamic} and \eqref{zeta2anotherdynamic} with $x_1$, $x_2$, $e$ being replaced by their steady-state values identified as $x_1^{s.s.}=\frac{e_0\zeta_2^2}{1+\zeta_2^2(1+\zeta_1)}$, $x_2^{s.s.}=\frac{e_0\zeta_1\zeta_2^2}{1+\zeta_2^2(1+\zeta_1)}$, and $e^{s. s.}=\frac{e_0}{1+\zeta_2^2(1+\zeta_1)}$, respectively.
	
	Thus, we arrive at the following rate equations of relative concentrations of substrate and product,
	\begin{subequations}
		\begin{align}
			\frac{d\zeta_1}{ d\theta}&=z-\frac{\zeta_1\zeta_2^2}{1+\zeta_2^2(1+\zeta_1)},\label{zeta1dynamic}\\
			\frac{d\zeta_2}{d \theta}&=\alpha_2[\frac{\zeta_1\zeta_2^2}{1+\zeta_2^2(1+\zeta_1)}-X_2\zeta_2]. 
			\label{zeta2dynamic}		
		\end{align}
	\end{subequations}
	Finally, introducing rescaled quantities $\tau=X_2^2z^{-2}\theta$, $x=X_2^{-1}z\alpha_2\zeta_1$, $y=X_2^{-1}z\zeta_2$, $\nu=X_2^{-3}z^4\alpha_2$, $\omega=X_2^{-1}z^2\alpha_2$, $\kappa=z{X_2}^{-1} \alpha_2$ in eq. \eqref{zeta1dynamic} and  \eqref{zeta2dynamic}, and taking a slow glycolytic flux limit during self-oscillation, we obtain the simplified form of the Selkov model system as, 
	\begin{equation}
		\begin{aligned}
		\frac{dx}{d\tau}&=\nu-xy^2 \\
			\frac{dy}{d\tau}&=xy^2-\omega y. 
			\label{reselkov}
		\end{aligned}
	\end{equation}
	
	The product formation step corresponding to $\rho=2$ in eq. \eqref{crn} of the Selkov system in sec. \ref{sec:1} is  irreversible. However, very often, the description of the chemical reaction network (e.g., nonequilibrium thermodynamic description) requires a chemical reaction network comprising all reversible elementary chemical reactions. Hence, we resort to an equivalent, completely reversible model\cite{reversibleselkoov} for more general convenience. This reversible version of the Selkov model includes the following chemical reactions, 
	
	\begin{equation}
		\begin{aligned}
			\rho&=1:&\ch{A&<=>[\text{k\textsubscript{1}}][\text{k\textsubscript{-1}}]S} \\
			\rho&=2:&\ch{S + 2 P &<=>[\text{k\textsubscript{2}}][\text{k\textsubscript{-2}}] 3 P}\\
			\rho&=3:& \ch{P&<=>[\text{k\textsubscript{3}}][\text{k\textsubscript{-3}}]
				B}\\ 
		\end{aligned}
		\label{revcrn}
	\end{equation}
	with S and P being the ATP and ADP, respectively. To serve our purpose of analysis, we have divided the chemical species of the reaction network into eq. \eqref{revcrn} into two sets: intermediate species, $I$ having dynamic concentration, and chemostatted species, $C$ with externally controllable concentration. Here $\{S, P\}\in I$ are intermediate species  $\{A, B\}\in C$ are chemostatted species. Assuming all the reverse rate constants $k_{-\rho}$ are vanishingly small($10^{-4}$), and the forward reaction rate constants $k_{\rho}$ are much higher relative to the reverse one, i.e., $k_{\rho}\gg k_{-\rho}$, we express the dynamical equation of eq. \eqref{revcrn} as 
	\begin{equation}
		\begin{aligned}
			\frac{ds}{dt}&=k_1a-k_2 s p^2\\
			\frac{dp}{dt}&=k_2s p^2-k_3p
			\label{revdynamic}
		\end{aligned}
	\end{equation}
	Considering scaled variables, $\tau=\frac{k_2t}{c_1^2}, \nu=\frac{c_1^3k_1a}{k_2},  x=c_1s, y=c_1p, \omega=\frac{c_1^2k_3}{k_2}$ in eq. \eqref{revdynamic} with $c_1$ being an arbitrary constant, we arrive at the same system as the eq.~\eqref{reselkov}. For convenience, we would use the notation $t$ for time throughout this report. 
	\section{\label{LSA}Reaction-diffusion Model of Glycolysis}
	We have Hopf instability\cite{strogatznonlinear}, characterized by uniform oscillation in the glycolysis model without any diffusive phenomena. From the eq. \eqref{reselkov}, the unique steady-state value of the system is obtained as, $x_0=\frac{\omega^2}{\nu}, y_0=\frac{\nu}{\omega}$. From the linear stability analysis at the steady-state value $(x_0,y_0)$, we can find the critical value of the control parameter for the onset of Hopf instability as $\nu_{cH}=\omega\sqrt{\omega}$. Further, the critical frequency of the Hopf instability is $f_{cH}=\omega$, and hence the approximate limit cycle period near the onset of instability, $\nu_{cH}$ is, $T=\frac{2\pi}{f_{cH}}$. 
	
	In a more general case, by taking the diffusion of the species into account, the generation of traveling waves having a nonzero finite wavenumber can be demonstrated in the model of glycolysis. The Selkov model can be represented as the reaction-diffusion equation in the presence of self diffusion. From eq. \eqref{reselkov}, the reaction-diffusion form of the Selkov model in one spatial dimension $r\in [0,l]$ reads
	\begin{equation} 
		\begin{aligned}
			\frac{\partial x}{\partial \tau}&=\nu-xy^2+D_{11}x_{rr} \\
			\frac{\partial y}{\partial \tau}&=xy^2-\omega y+D_{22}y_{rr} 
			\label{ddynamic}
		\end{aligned}
	\end{equation}
	with $D_{11}$ , $D_{22}$ being constant self-diffusion coefficients of intermediate species $x$ and $y$, respectively. 
	We would express the concentration evolution of the system \eqref{ddynamic} analytically after introducing the amplitude equation representation in the next section.
	
	In the presence of diffusion, the critical value of the control parameter, $\nu$, for the onset of the traveling waves is identified as,
	\begin{equation}
		\nu_{ctw}=w\sqrt{w-(D_{11}+D_{22})q^2}. 
	\end{equation} 
	According to the zero flux boundary conditions, the wavenumber $q$ of the traveling wave follows $q=\frac{n\pi}{l}$ with $n$ being an integer. Additionally, the traveling wave with wavenumber $q$ in this model satisfies the condition, $D_{22}^2q^4+2D_{11}wq^2-w^2\leq0.$  
	
	\section{\label{selampli}Amplitude Equation Representation of  the Reaction-diffusion Glycolysis Model}
	To encapsulate the role of nonlinearity in the evolution of the dynamic entity of the system, it is quite standard practice to investigate the amplitude dynamics of the system. In the presence of diffusion, the complex Ginzburg-Landau equation(CGLE)\cite{aranson2002world, Cross2009PatternSystems} is the lowest order equation that can properly capture the amplitude of the spatially extended nonlinear system near the onset of Hopf instability. For a nonlinear reaction-diffusion model, the CGLE can be expressed as,
	\begin{equation}
		\frac{\partial Z}{\partial t}=\lambda Z -(\beta_{\mathcal{R}}-i\beta_{\mathcal{I}})\mid \color{black}Z \mid ^2 Z+(\alpha_{\mathcal{R}}+ i \alpha_{\mathcal{I}})\color{black}\partial_{r}^2 Z,
		\label{cgl}
	\end{equation}
	with $Z$ being the complex amplitude and $\lambda$, $\beta_{\mathcal{R}}$, $\beta_{\mathcal{I}}$, $\alpha_{\mathcal{R}}$, and $\alpha_{\mathcal{I}}$ being coefficients comprising the details of the system. In a comoving coordinate of $r=r-velocity\times t $, the traveling waves and related instabilities can be explored by employing the same amplitude equation as of eq. \eqref{cgl}. If we consider a complex amplitude of the form, $Z=\mathcal{A}\exp({i \phi})$ in eq. \eqref{cgl} and separate real and imaginary parts, two different dynamical equations for the magnitude and the phase would emerge, 
	\begin{subequations}
		\begin{align}
			\frac{\partial \mathcal{A}}{\partial t}&=\lambda \mathcal{A} {-\beta_{\mathcal{R}}\mathcal{A}^3-\alpha_{\mathcal{I}}(2\mathcal{A}_{r}\phi_{r}+\phi_{rr}\mathcal{A})+\alpha_{\mathcal{R}}}(\mathcal{A}_{rr}-\mathcal{A}\phi_{r}^2), \label{ampd}\\
			\frac{\partial \phi}{\partial t}&={\beta_{\mathcal{I}}\mathcal{A}^2+\alpha_{\mathcal{R}}(\frac{2\mathcal{A}_{r}\phi_{r}}{\mathcal{A}}+\phi_{rr})+\alpha_{\mathcal{I}}}(\frac{\mathcal{A}_{rr}}{\mathcal{A}}-\phi_{r}^2). \label{phased}
		\end{align}
	\end{subequations} 
	
	Now we can represent the normal form\cite{Nicolis1995IntroductionScience, Cross2009PatternSystems} of CGLE in eq. \eqref{cgl} as 
	\begin{equation}
		\frac{\partial Z}{\partial t}=\lambda Z -(1-i\beta)\mid Z \mid ^2Z+(1+i \alpha)\partial_{r}^2 Z,
		\label{ncgle}
	\end{equation} 
	with coefficients in normal form are given as {$\alpha=\frac{\alpha_{\mathcal{I}}}{\alpha_{\mathcal{R}}}$ and $\beta=\frac{\beta_{\mathcal{I}}}{\beta_{\mathcal{R}}}.$} To obtain the coefficients, $\alpha$ and $\beta$, we need to acquire the magnitude and phase dynamical equations from eq. \eqref{ddynamic} representing the spatially extended Selkov model in the presence of diffusion. For this purpose, we have implemented the Krylov-Bogoliubov(KB) averaging method\cite{krylov1949introduction}. Then comparing the amplitude equation deduced by KB averaging method with eq. \eqref{ampd} and \eqref{phased}, we arrive at the coefficients of the form, $\alpha=\frac{\Omega(D_{22}-D_{11})}{(D_{11}+D_{22})}$ and $\beta=-\frac{p_2}{p_1}\frac{\sqrt{\omega}\omega}{3\nu }$, with $\Omega=\frac{\nu}{\sqrt{\omega}}$ and $\frac{p_2}{p_1}=\frac{2c_1^2}{\nu c_2^2}$. Terms $c_1=\frac{(2\omega^{-2}\nu-\frac{\omega}{\nu})}{2\lambda}, c_2=\frac{\omega^{-2}}{2\lambda}$ are coined in the KB averaging method and correction factors $p_1$ and $p_2$ are introduced in KB amplitude and phase equation to properly capture the radius and phase modification owing to unidirectional acceleration from unstable steady state\cite{Lavrova2009PhaseInflux}. The details of deriving the phase and amplitude equation using the KB averaging method for a more general reaction-diffusion equation with both self- and cross-diffusion can be found in ref. \cite{pkgg,pkgg2}.   
	
	\subsection{\label{sec:hae}Amplitude and Phase of the System}   
	For the normal form of the CGLE eq. \eqref{ncgle}, we can represent eq. \eqref{ampd} and \eqref{phased} as 
	\begin{subequations}
		\begin{align}
			\frac{\partial \mathcal{A}}{\partial t}&=\lambda \mathcal{A} -\mathcal{A}^3-\alpha(2\mathcal{A}_{r}\phi_{r}+\phi_{rr}\mathcal{A}) +(\mathcal{A}_{rr}-\mathcal{A}\phi_{r}^2), \label{samp}\\
			\frac{\partial \phi}{\partial t}&=\beta \mathcal{A}^2+(\frac{2\mathcal{A}_{r}\phi_{r}}{\mathcal{A}}+\phi_{rr})+\alpha(\frac{\mathcal{A}_{rr}}{\mathcal{A}}-\phi_{r}^2). \label{sphase}
		\end{align}
	\end{subequations}
	Assuming slow temporal variation of amplitude and long-range phase variation, we can write the (steady state) amplitude from eq. \eqref{samp} as,  
	\begin{equation}
		\mathcal{A}^2=\lambda -\alpha \phi_{rr} -\phi_{r}^2. \label{overssamp}
	\end{equation}
	With these considerations and amplitude variation in eq. \eqref{overssamp}, the following phase dynamics can be derived from eq. \eqref{sphase}
	\begin{equation}
		\frac{\partial \phi}{\partial t}=\beta\lambda +(1-\alpha\beta)\phi_{rr}-(\alpha+\beta)\phi_{r}^2. \label{nonlinearphase}	
	\end{equation}	
	In the above equation, $(\alpha+\beta)<0$, and $(1-\alpha\beta)<0$
	set criterion of inward and outward rotating spiral exchange and the Benjamin-Feir(BF) instability onset~\cite{benjamin1967instability, Cross2009PatternSystems}, respectively. Furthermore, solving eq.~\eqref{nonlinearphase}, we have the closed form of the system phase~\cite{pkgg2} as
	\begin{equation}
		\phi=\beta \lambda t-\frac{1-\alpha \beta}{\alpha+\beta}[\ln{G_0}+(1-\alpha \beta)(\alpha+\beta)^2q^2t+(\alpha+\beta)qr],
		\label{nonlinearphaseeq}
	\end{equation} 
	with $G_0$ being an arbitrary initial value.
	
	\subsection{Lambda–omega Representation}
	In this investigation, we would consider equal diffusivities in the reaction-diffusion form of the Selkov model(eq. \eqref{ddynamic}). Hence, we can also represent oscillatory reaction-diffusion equations in terms of prototype ``Lambda–omega" systems\cite{lamomega1}:
	
	\begin{equation} 
		\begin{aligned}
			\frac{\partial x}{\partial t}&=\Lambda(\mathcal{A})x-\tilde{\Omega}(\mathcal{A})y+x_{rr} \\
			\frac{\partial y}{\partial t}&=\tilde{\Omega}(\mathcal{A})x+\Lambda(\mathcal{A})y+y_{rr},
			\label{lambdaomega}
		\end{aligned}
	\end{equation}
	where $\Lambda$ and $\tilde{\Omega}$ are real functions of variable $\mathcal{A}$ with $\mathcal{A}^2=x^2+y^2$, and $r$ is a scaled variable. Now consideration of complex entity $Z=x+iy=\mathcal{A}\exp({i \phi})$  aids us to write the following equation from eq. \eqref{lambdaomega},
	\begin{equation}
		\frac{\partial Z}{\partial t}=(\Lambda(\mathcal{A})+i\tilde{\Omega}(\mathcal{A}))Z+\partial_{r}^2 Z.\label{comlamomega}
	\end{equation} 
	For equal diffusion coefficients, the CGLE of eq. \eqref{ncgle} reads,
	\begin{equation}
		\frac{\partial Z}{\partial t}=(\lambda -(1-i\beta)\mid Z \mid ^2)Z+\partial_{r}^2 Z,
		\label{rencgle}
	\end{equation} 
	as $\alpha$ vanishes. In eq. \eqref{comlamomega} and \eqref{rencgle}, complex-valued entity $Z=\mathcal{A}\exp({i \phi})$ and both equations renders traveling waves here under the equal self-diffusivities in the reaction-diffusion system and hence both of these equations can be treated on an equal footing. So comparing eq. \eqref{comlamomega} and \eqref{rencgle}, we obtain the functions, $\Lambda$ and $\tilde{\Omega}$ as  
	\begin{equation} 
		\begin{aligned}
			\Lambda(\mathcal{A})=\lambda-\mathcal{A}^2\\
			\tilde{\Omega}(\mathcal{A})=\beta \mathcal{A}^2\label{lamomegaform}.
		\end{aligned}
	\end{equation}	
	When the CGLE in eq. \eqref{rencgle} admits the asymptotic plane wave solution of the form $Z=\mathcal{A}\exp{i(\omega_{0}t+qr)}$ with $\omega_{0}$ being the shift in frequency from the critical frequency $f_{cH}$, and $q$ being a unique wavenumber selected by the unique spiral frequency, we can express
	$q^2=\lambda-\mathcal{A}^2$, $\omega_{0}=\beta(\lambda-q^2)=\beta\mathcal{A}^2$.
	Taking $q=0$ in $\omega_{0}$ expression results in the bulk frequency for the system. Further, surveying eq. \eqref{lamomegaform}, we gather $q^2=\Lambda(\mathcal{A}),$ and $\omega_{0}=\tilde{\Omega}(\mathcal{A}).$
	\subsection{Plane Waves Stability and Wavenumber Selection}
	The wavenumber selection in a finite system far from equilibrium is a strenuous job due to the complex dependence of the wavenumber on the control parameter values, boundary conditions, dynamical processes, perturbations, and methodologies. In this regard, we can investigate the stability of the asymptotic plane wave by taking a perturbation about the nonlinear wave state. In this way, the threshold of stable  wavenumber above which the traveling wave becomes unstable can be found as
	\begin{equation}
		q^2= \frac{\lambda}{2\beta^2+3}
		\label{selected_wavenumber}
	\end{equation} 
	provided $\alpha=0$ in the case of equal self-diffusivities. Thus,  eq. \eqref{selected_wavenumber} aids us in identifying the band of allowed wavenumber of the traveling wave. Further, it is evident from eq. \eqref{selected_wavenumber} that the allowed wavenumber, $q$, has explicit dependence on the amplitude equation coefficients $\beta$. 
	
	\section{Analytical Expression of Concentration Evolution}   
	The concentration evolution of species $x$ and $y$ for the traveling waves near the onset of oscillation can be expressed as
		\begin{equation}
			{\mathcal{Z}_I}_{TW}={\mathcal{Z}_I}_{0}+A_{TW}U_{cH}\exp{(i f_{cH}t)}+C.C.,
			\label{hwave}
		\end{equation}   
		with ${\mathcal{Z}_I}_{0}$ being the uniform base state and $A_{TW}$ being the amplitude part within the oscillatory regime obtained from eq. \eqref{ncgle}, and $U_{cH}$ is the critical eigenvector corresponding to the jacobian matrix of the linearized system. For traveling waves, the amplitude part $A_{TW}$ also has the spatial variation due to nonzero wavenumber, and the concentration dynamics within the oscillatory regime become
		\begin{widetext}
			\begin{gather}
				\begin{pmatrix}
					x\cr
					y
				\end{pmatrix}=
				\begin{pmatrix}
					x_0\cr
					y_0
				\end{pmatrix}
				+ \sqrt{\lambda-q^2}
				\begin{pmatrix}
					2\cos{(\omega_{0}t+f_{cH} t+qr)}+2\sin{(\omega_{0}t+f_{cH} t+qr)}\cr
					-2\cos{(\omega_{0}t+f_{cH} t+qr)}
				\end{pmatrix}.
				\label{maineq}
			\end{gather}
	\end{widetext}
	\section{\label{influx}Inhomogeneous Chemostatted Influx and Numerical Investigation}
	In contrast to ref.~\cite{pkgg2}, we have considered an inhomogeneous inflow of the scaled chemostatted species within the system instead of a uniform influx motivated by the modeling of glycolytic wave propagation by Lavrova et al. \cite{LAVROVA2009127, Lavrova2009PhaseInflux}. {In ref.~\cite{LAVROVA2009127, Lavrova2009PhaseInflux}, the parabolic inflow profile was introduced by realizing the experimental situation for glycolytic oscillations and waves in an open spatial reactor~\cite{BAGYAN200567, spatialreactor}. A semipermeable membrane separates a gel layer from the reservoir in the open spatial reactor. The glycolytic reaction happens at this gel layer. The inflow of chemostatted species through this gel layer can be theoretically perceived by considering a parabolic inflow profile as a first-order correction to the homogeneous influx profile.} To keep this inhomogeneous inflow profile in accord with the percolation of species in the experimental setting of an open spatial reactor, a  vertex form of the parabola is considered,   
	\begin{equation}
		\nu(r)=\nu_0+\frac{(\nu_b-\nu_0)}{r_0^2}(r-r_0)^2,
		\label{inhomoflux}
	\end{equation}
	where the vertex of the parabola is $(r_0,\nu_0)$ and $\nu_b$ is the chemostatted species concentration at the borders. Due to this inhomogeneous consideration of the $\nu$, the steady state value of the system $(x_0,y_0)$ now has a nonuniform profile. For a constant value of rate constants of the chemical reaction network, it is apparent from the discussion on the reversible Selkov model in sec. \ref{sec:1} that inhomogeneous inflow of the scaled chemostatted species $\nu$ is owing to the inhomogeneous inflow of the chemostatted species $A$. This spatial dependence of the $\nu$ modifies the amplitude equation coefficient, $\beta$, and thus the profile of the phase and band of allowed wavenumbers can be modified.   
	
	For the inhomogeneous $\nu$, we have numerically integrated the reaction-diffusion equation in eq. \eqref{ddynamic} using the Python programming language(version 3.9) and the `LSODA’~\cite{LSODA} integration method of the scipy.integrate.solve\_ivp class of the SciPy library(version 1.8). For the `LSODA’, specifying `lband' and `uband' parameters regarding the bandwidth of the jacobian would significantly speed up the computation here. For numerical work, second-order spatial derivatives are represented in terms of 1D  discrete Laplace operators obtained by a finite differences method. We have uniformly divided the one-dimensional spatial length of $l=1$ into 300 grid points. Our simulation has been performed in a one-dimensional spatial domain with no-flux boundary conditions,
	$\frac{\partial x}{\partial r}(0,\tau)=\frac{\partial x}{\partial r}(l,\tau)=0,\frac{\partial y}{\partial r}(0,\tau)=\frac{\partial y}{\partial r}(l,\tau)=0.$ For numerical integration, we have used a time-step size of $\Delta {\tau}=\frac{0.3h^2}{D_{11}}$ with $h$ being the size of the space step. For numerical simulation, the system is initially kept at a state very close to the steady state of the system.

	\section{\label{reldiss}Results and Discussion}
	
	\begin{figure}
		\begin{center}
			\includegraphics[width=\linewidth]{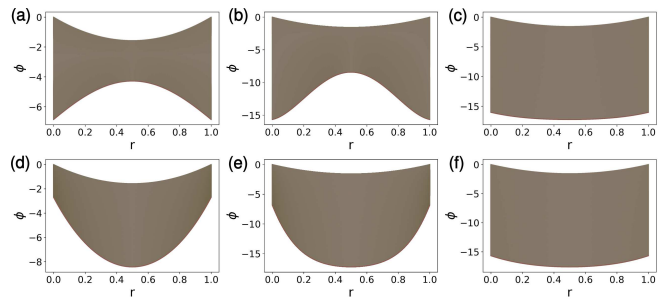}
			\caption{\label{fig1} Spatial variations of the phase, $\phi$ (without taking wavenumber into account) for different inhomogeneous influx profiles are illustrated for the time range of $t=0$ to $t=75$. In upper panel, (a)$\nu_0=2.80$ and $\nu_b=2.75$, (b)$\nu_0=2.75$ and $\nu_b=2.55$, (c)$\nu_0=2.60$ and $\nu_b=2.55$. In lower panel, values of $\nu_0$ and $\nu_b$ in (a), (b), and (c) are interchanged and thus flip the inhomogeneous profile, i.e.,  (d)$\nu_0=2.75$ and $\nu_b=2.8$, (e)$\nu_0=2.55$ and $\nu_b=2.75$, (f)$\nu_0=2.55$ and $\nu_b=2.60$. Initial phase is taken as $\phi(r,0)=2\pi r(r-1)$ in all cases.}
		\end{center}
	\end{figure}
	\begin{figure}
		\begin{center}
			\includegraphics[width=\linewidth]{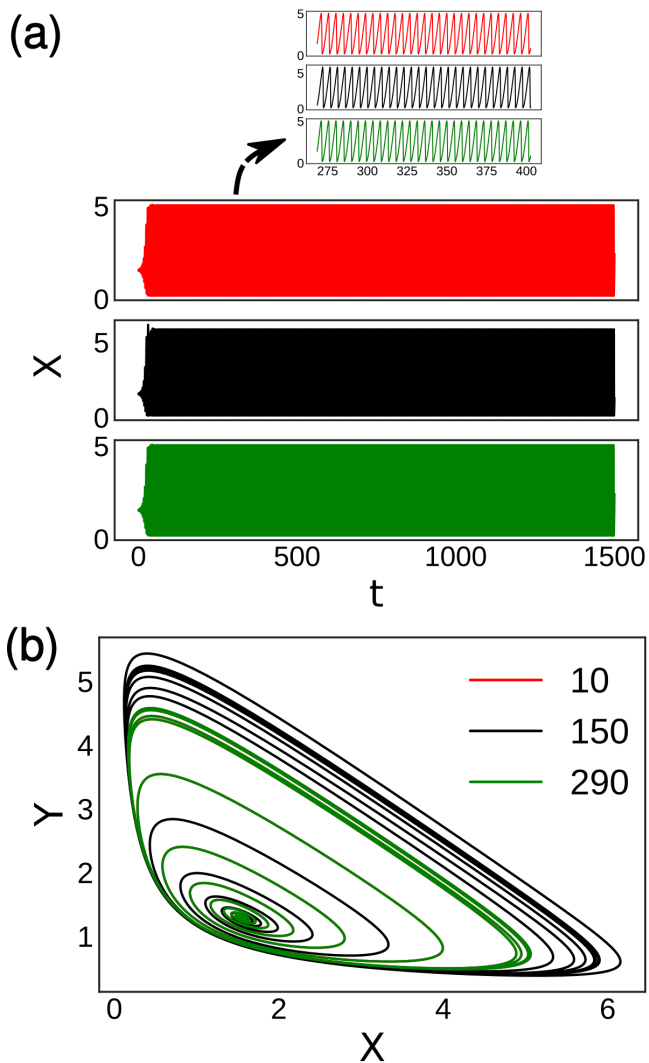}
			\caption{\label{fig2} Concentration profiles of intermediate species, $X$ for symmetric parabolic influx with $\nu_0=2.55$ and $\nu_b=2.60$. Relaxation oscillation at the center and two boundaries scenario. (a)Temporal dynamics of concentration near left boundary, $r=10$(red), at center, $r=150$(black), and near right boundary, $r=290$(green). (b)Phase portraits at the same spatial regimes. For diffusion coefficients, $D_{11}=D_{22}=0.0025$ periodic oscillation is shown as a subplot for the small time window.}
		\end{center}
	\end{figure}
	
	\begin{figure*}
		\begin{center}
			\includegraphics[width=\textwidth]{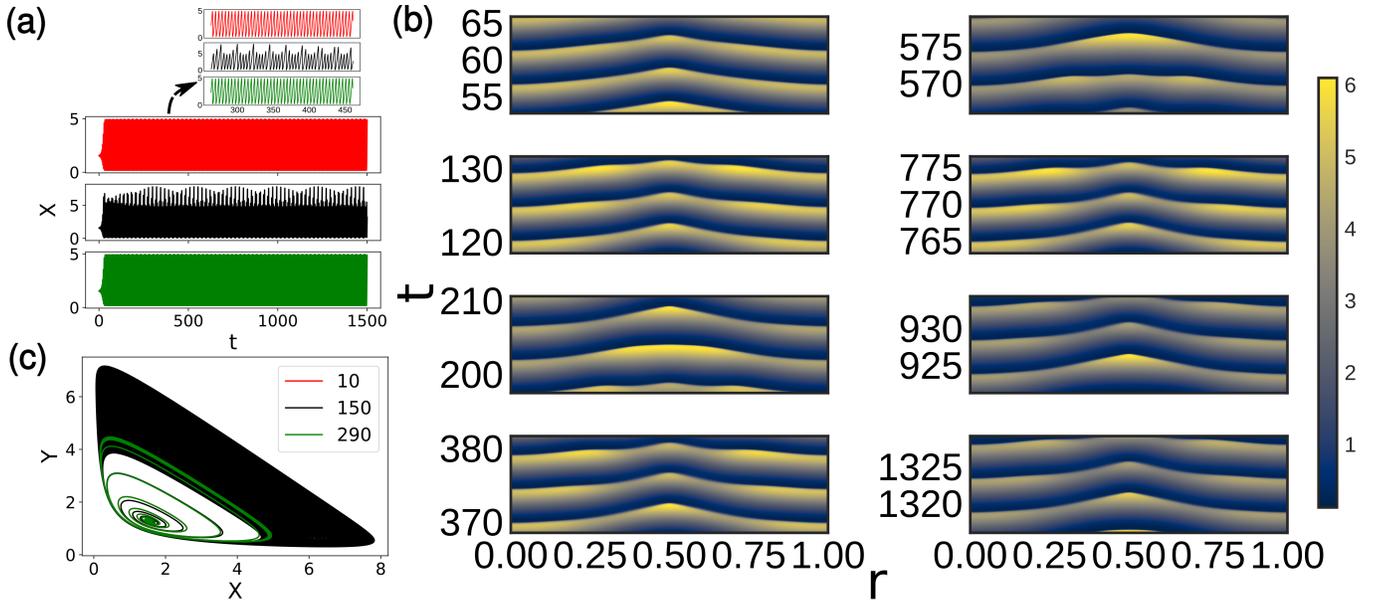}
			\caption{\label{fig3} Concentration profiles of intermediate species, $X$ for symmetric parabolic influx with $\nu_0=2.55$ and $\nu_b=2.60$ are illustrated for diffusion coefficients, $D_{11}=D_{22}=0.00051$. (a)Concentration dynamics reveal that the central regime develops a quasiperiodic behavior over the whole time range. (b)Spatiotemporal images are shown at different time intervals. No wave direction change is spotted. (c)The phase portrait at the central regime demonstrates dense nature. }
		\end{center}
	\end{figure*}

	\begin{figure*}
		\begin{center}
			\includegraphics[width=\textwidth]{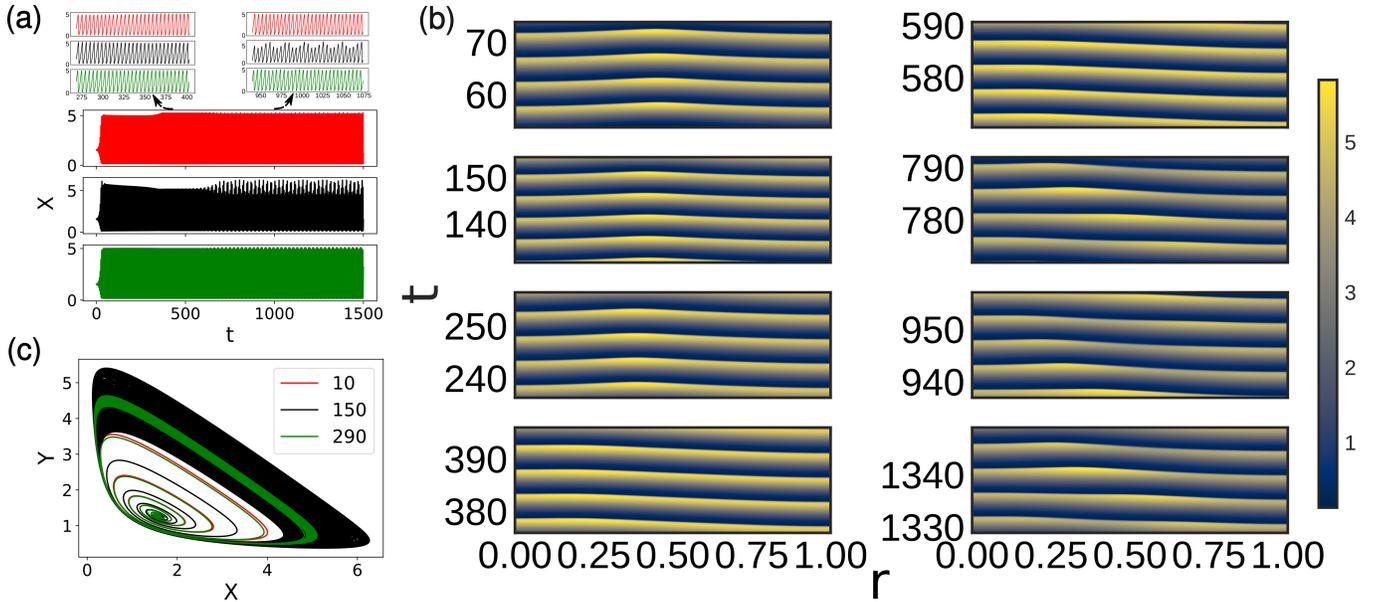}
			\caption{\label{fig4} Concentration profiles of intermediate species, $X$ for slightly asymmetric parabolic influx characterized by $r_0=0.495$ are shown for $\nu_0=2.55$ and $\nu_b=2.60$ with diffusion coefficients, $D_{11}=D_{22}=0.0025$. (a)Concentration dynamics of the center regime exhibit a transition from periodic to complex oscillation after a certain time. (b)Spatiotemporal images are shown at different time intervals. Wave direction change is noted in the presence of weak asymmetry. (c)The phase portrait at the center regime demonstrates dense nature for a higher time range.}
		\end{center}
	\end{figure*}  
	\begin{figure*}
		\begin{center}
			\includegraphics[width=\textwidth]{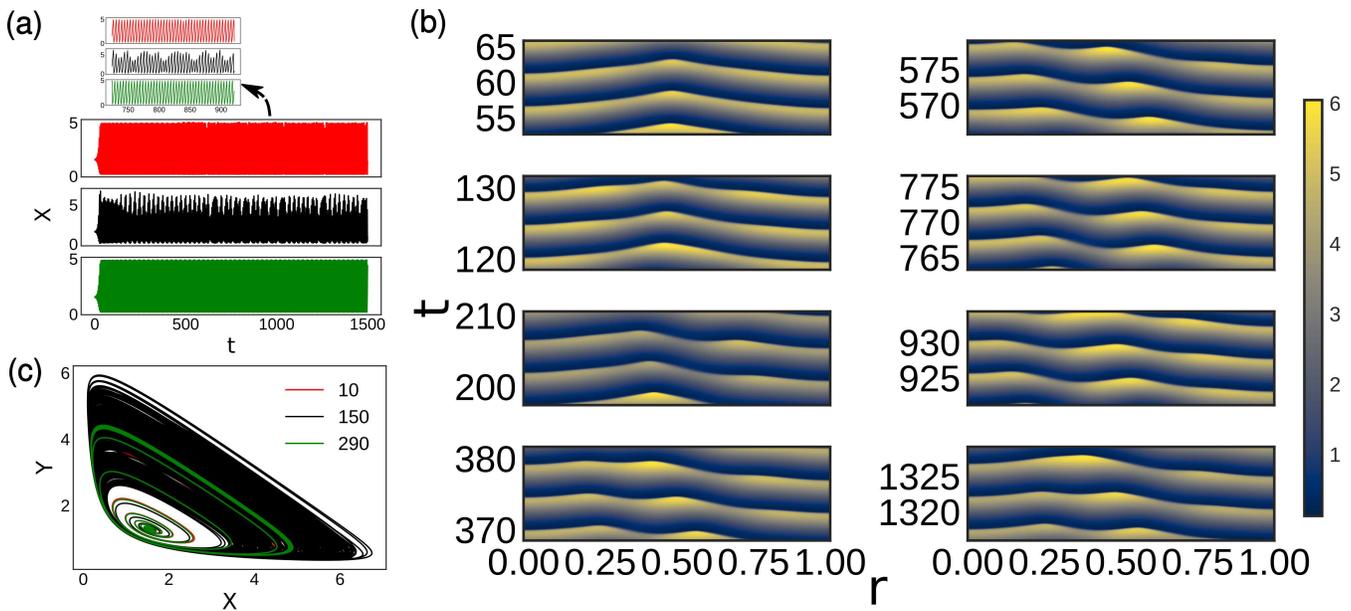}
			\caption{\label{fig5} Concentration dynamics of $X$ for slightly asymmetric parabolic influx($r_0=0.495$) are obtained for $\nu_0=2.55$ and $\nu_b=2.60$ with diffusion coefficients, $D_{11}=D_{22}=0.00051$. (a)The center regime demonstrates a chaotic oscillation, while near the left boundary, we have a certain drop in the periodic oscillation. (b)Spatiotemporal images capture the wave direction change. (c) Phase portraits corresponding to all three regions are shown.}
		\end{center}
	\end{figure*}
	\begin{figure*}
		\begin{center}
			\includegraphics[width=\textwidth]{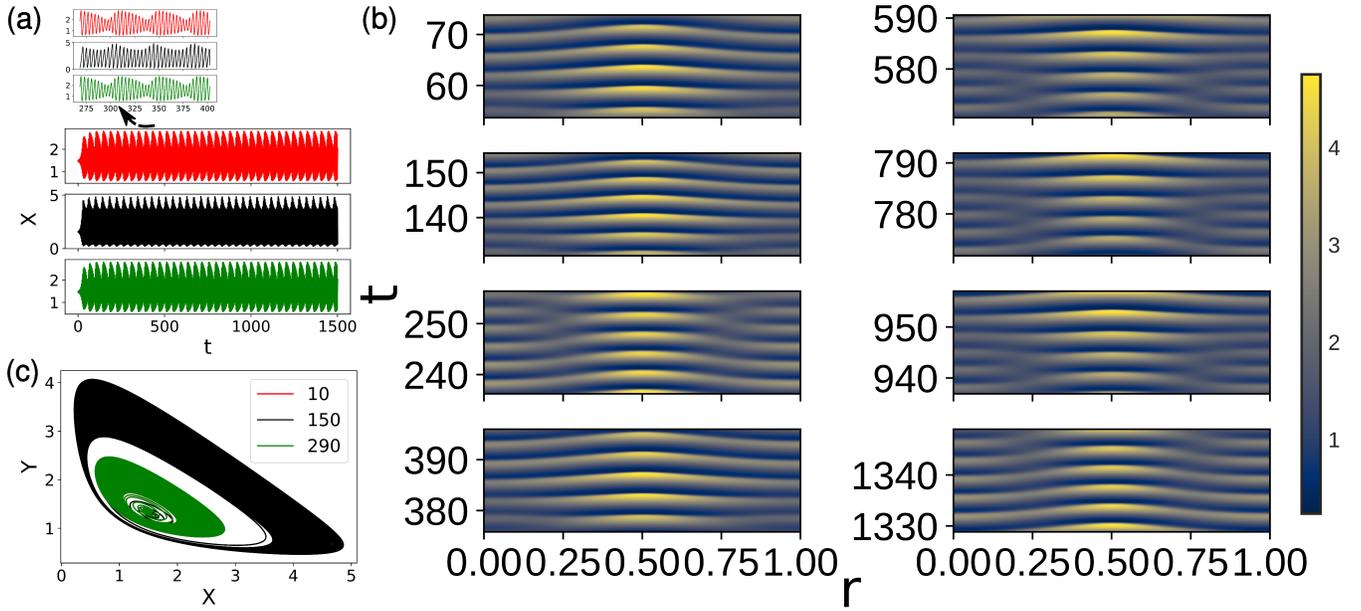}
			\caption{\label{fig6} Here $\nu_0=2.55$ and $\nu_b=2.75$ in symmetric influx profile. (a) Amplitude-modulated concentration profiles for all three regimes are acquired for $D_{11}=D_{22}=0.0025$. (b)Spatiotemporal images have different wavenumbers near borders and the center. (c)Dense phase portrait for both the center and border regimes.}
		\end{center}
	\end{figure*} 
	
	\begin{figure*}
		\begin{center}
			\includegraphics[width=\textwidth]{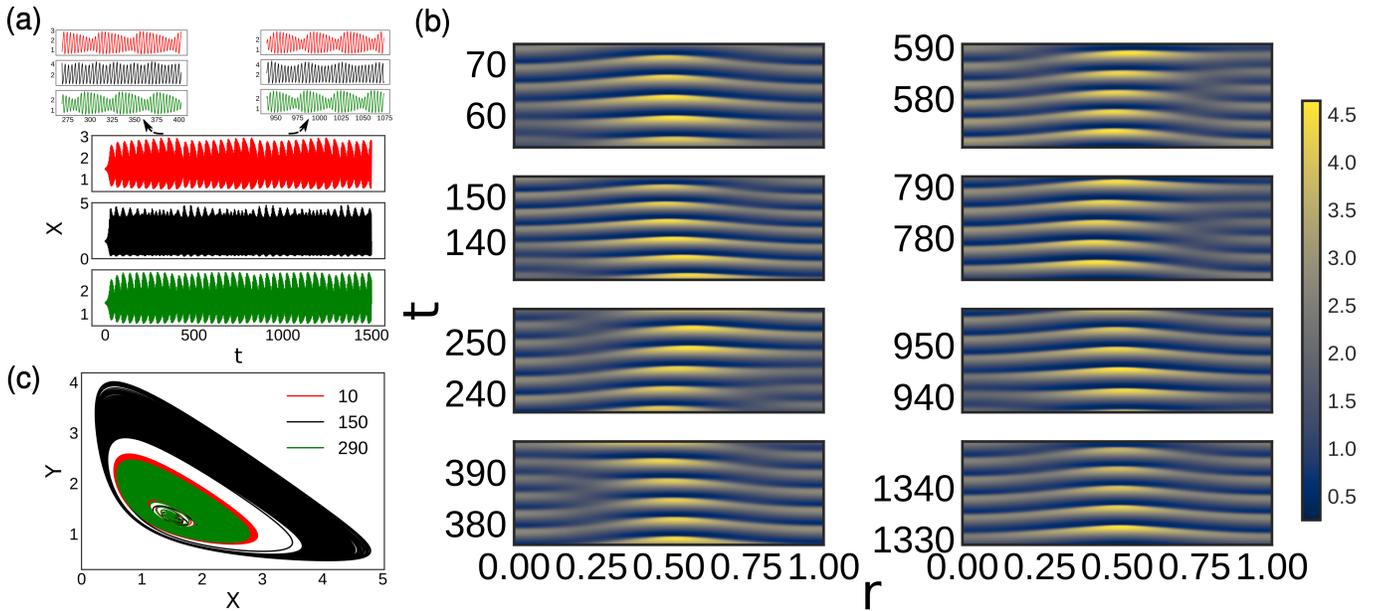}
			\caption{\label{fig7} (a)Amplitude modulation in the concentration dynamics of all three regimes for $\nu_0=2.55$ and $\nu_b=2.75$, and diffusion coefficients, $D_{11}=D_{22}=0.0025$. (b)Spatiotemporal images capture sequences of wave direction switch in the presence of slight asymmetry($r_0=0.495$). (c)Corresponding phase portraits are illustrated.}
		\end{center}
	\end{figure*}
	\begin{figure*}
		\begin{center}
			\includegraphics[width=\textwidth]{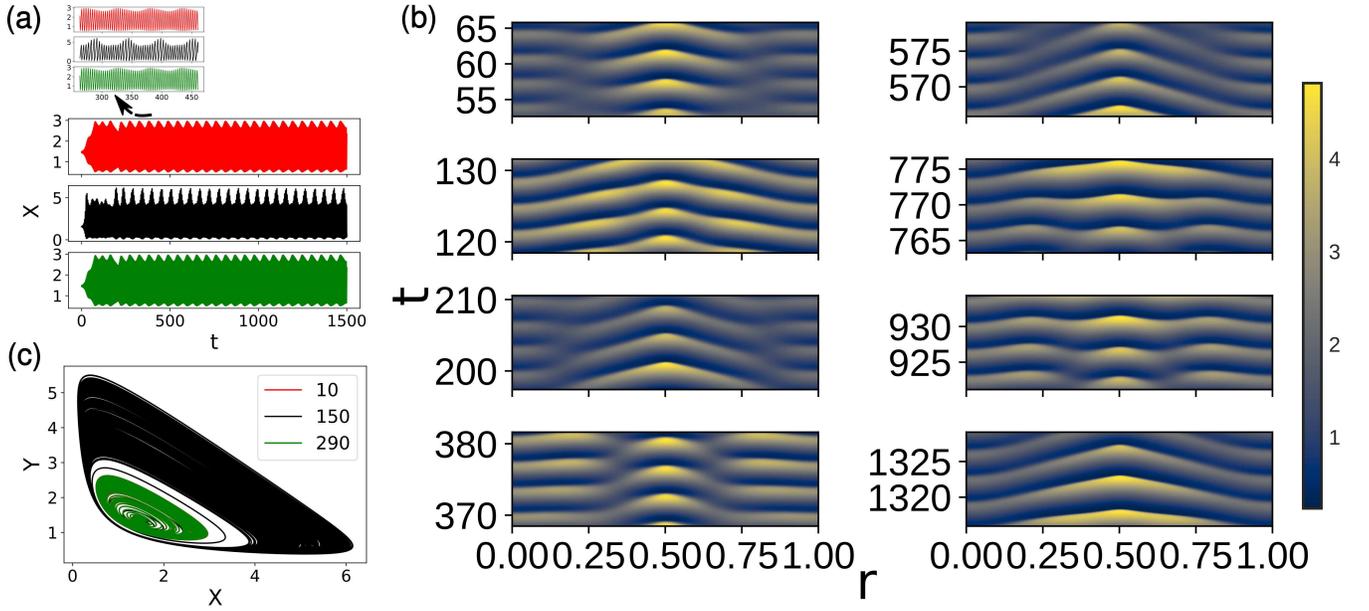}
			\caption{\label{fig8} (a)For symmetric influx profiles $\nu_0=2.55$ and $\nu_b=2.75$, modulation of the concentration profiles at the center regimes are stronger than the boundaries for low diffusion magnitude, $D_{11}=D_{22}=0.00051$. (b)Spatiotemporal images at different time intervals exhibit multifold wave profiles but no change in wave direction. (c)Corresponding phase portraits are illustrated.}
		\end{center}
	\end{figure*} 
	\begin{figure*}
		\begin{center}
			\includegraphics[width=\textwidth]{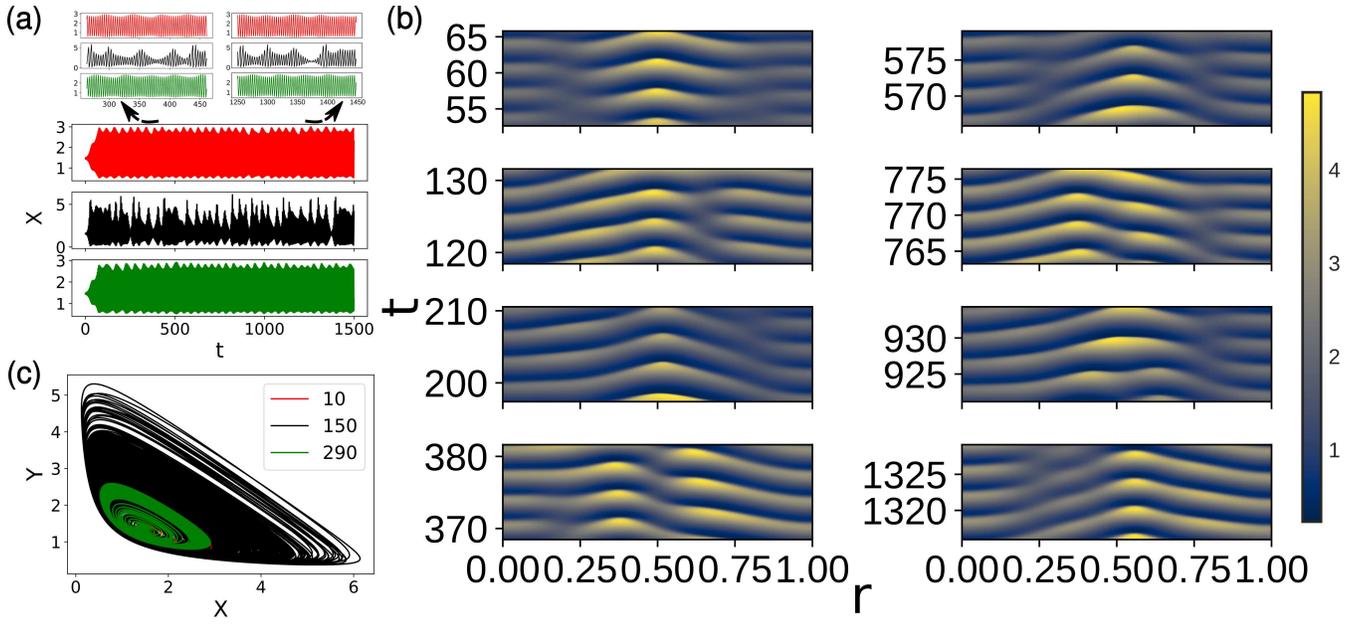}
			\caption{\label{fig9} For $\nu_0=2.55$ and $\nu_b=2.75$, asymmetry in the influx profile($r_0=0.495$) yields a quasiperiodic route to chaos in concentration of $X$ at the center regime.  $D_{11}=D_{22}=0.00051$. (b)Spatiotemporal images demonstrate a slight shift in the wave propagation direction (c)The phase portrait also reflects the chaotic dynamics in the center regime.}
		\end{center}
	\end{figure*} 	
	\begin{figure*}
		\begin{center}
			\includegraphics[width=\textwidth]{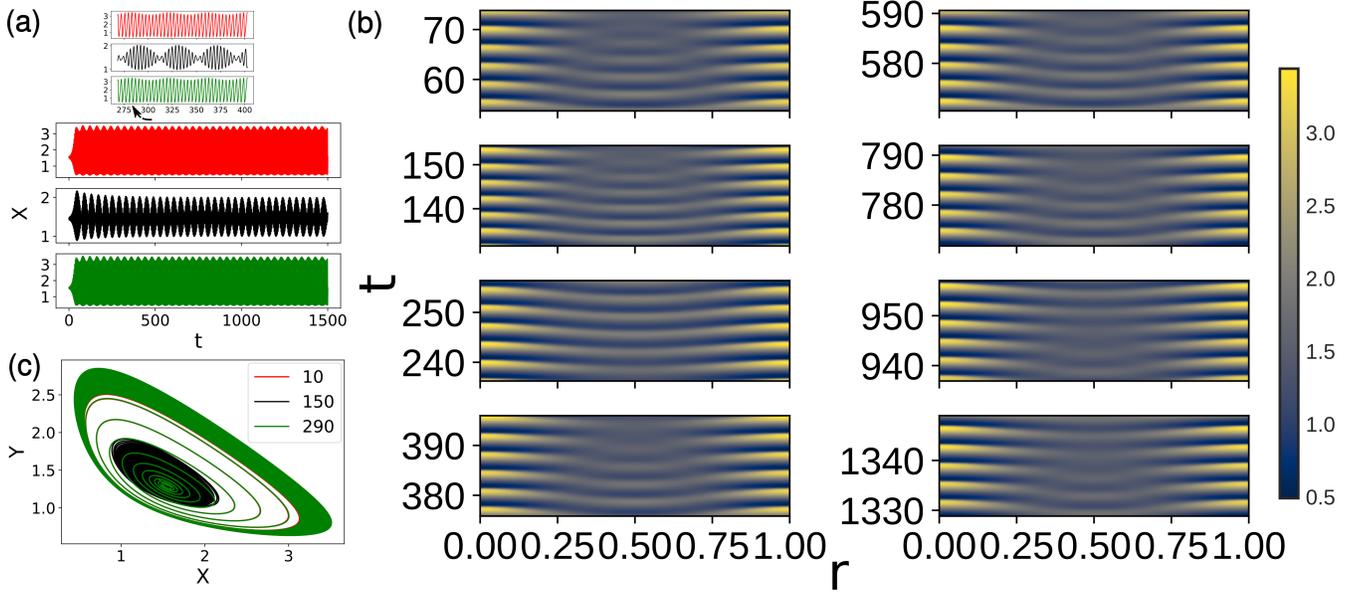}
			\caption{\label{fig10} (a)For $D_{11}=D_{22}=0.0025$, the concentration dynamics at the center regime exhibit a beat interference pattern for a symmetric influx profile with $\nu_0=2.75$ and $\nu_b=2.55$. (b)Spatiotemporal images illustrate outwardly moving traveling waves. (c)The corresponding phase portraits are shown.   }
		\end{center}
	\end{figure*} 
	\begin{figure*}
		\begin{center}
			\includegraphics[width=\textwidth]{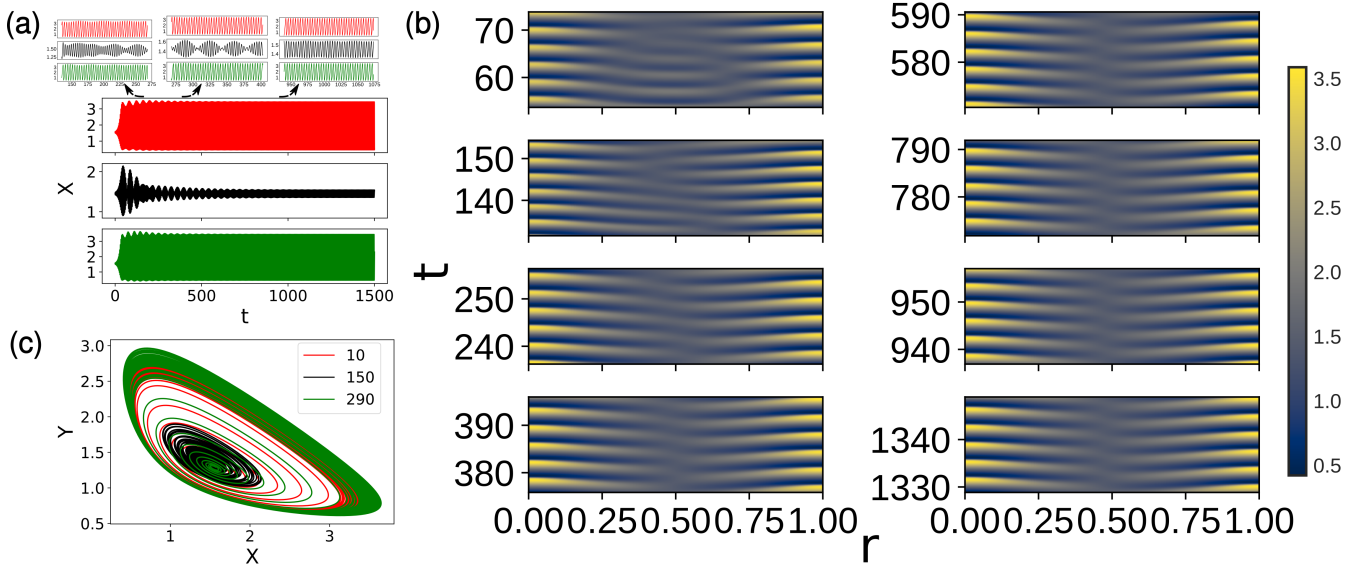}
			\caption{\label{fig11} (a)For asymmetric influx with $\nu_0=2.75$ and $\nu_b=2.55$, the concentration dynamics at the center regime exhibit a dynamic transition from quasiperiodic to periodic pattern in the presence of of diffusion coefficients, $D_{11}=D_{22}=0.0025.$ (b)Spatiotemporal images suggest only outwardly moving traveling waves. (c)The phase portrait corresponding to the center regime indicates a dynamic transition.}
		\end{center}
	\end{figure*} 
	\begin{figure*}
		\begin{center}
			\includegraphics[width=\textwidth]{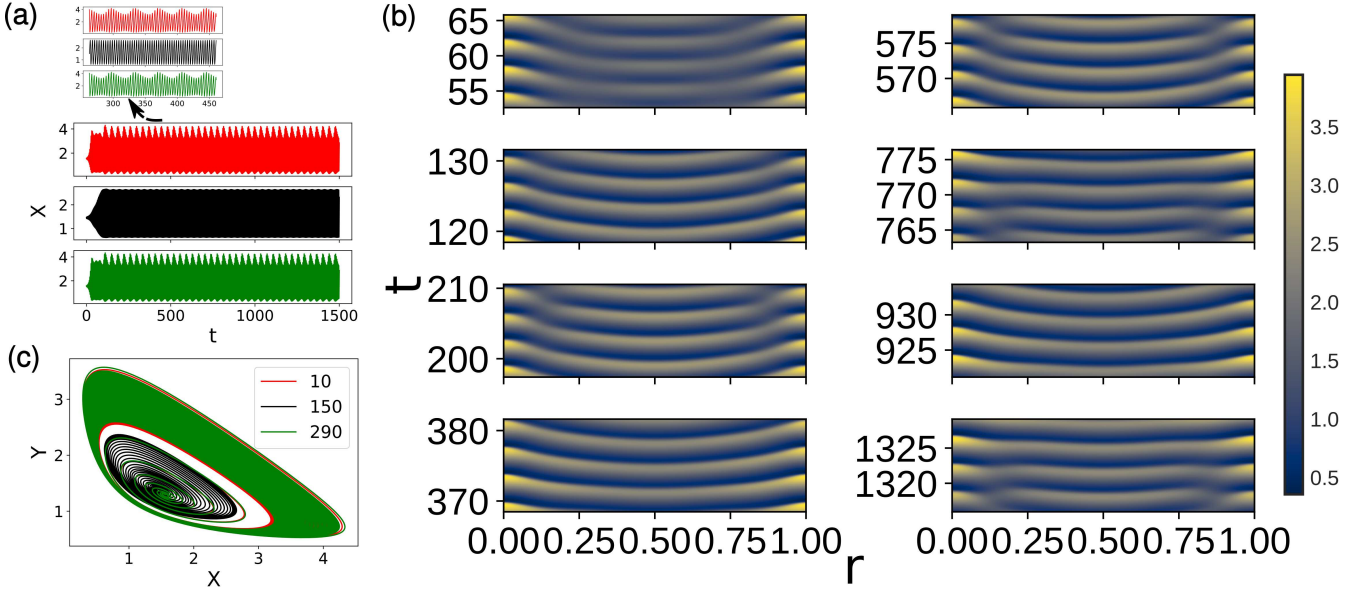}
			\caption{\label{fig12} For diffusion coefficients, $D_{11}=D_{22}=0.00051$ and symmetric influx with $\nu_0=2.75$ and $\nu_b=2.55$, (a)periodic behavior at the center regime and quasiperiodic pattern near two boundaries are found in the concentration dynamics of $X$. (b)Spatiotemporal images show that waves always move outwardly. (c)The corresponding phase portrait is illustrated.}
		\end{center}
	\end{figure*}
	\begin{figure*}
		\begin{center}
			\includegraphics[width=\textwidth]{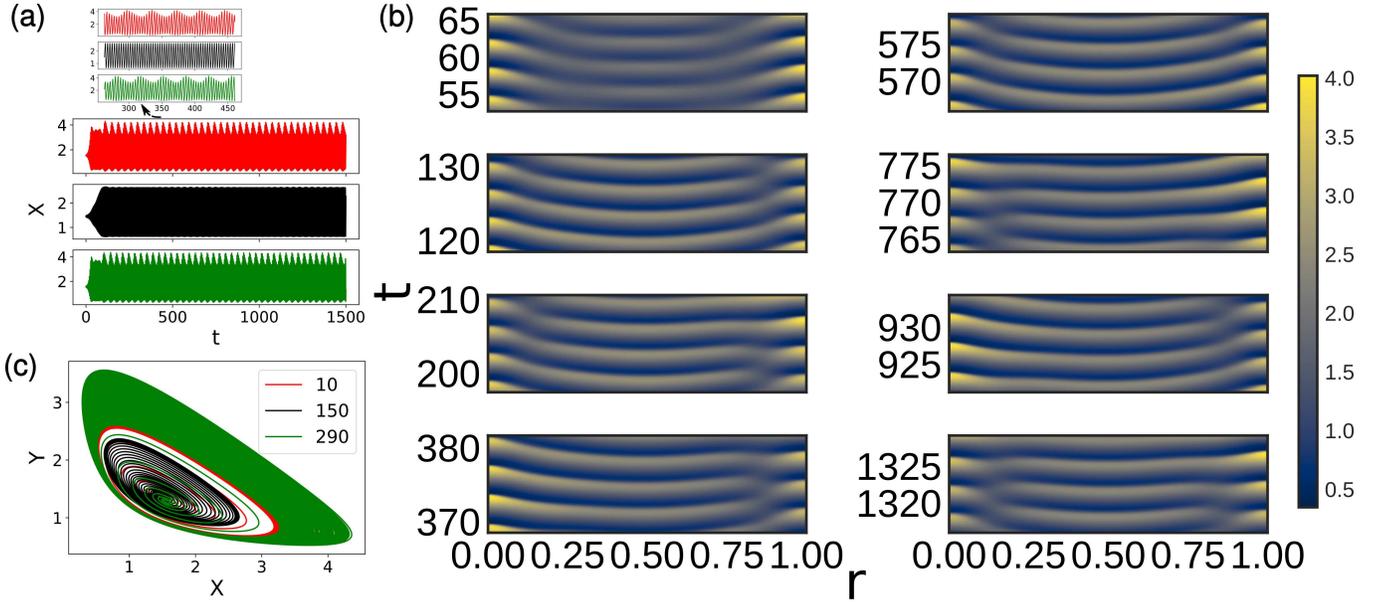}
			\caption{\label{fig13} For asymmetric influx with $\nu_0=2.75$ and $\nu_b=2.55$ and diffusion coefficients, $D_{11}=D_{22}=0.00051$, (a)Concentration dynamic behavior near the center regime and near two boundaries exhibit similar pattern as in the symmetric influx profile case. (b)Spatiotemporal images reveal a wave direction change. (c)A corresponding phase portrait is shown.}
		\end{center}
	\end{figure*} 	
	\begin{figure*}
		\begin{center}
			\includegraphics[width=\textwidth]{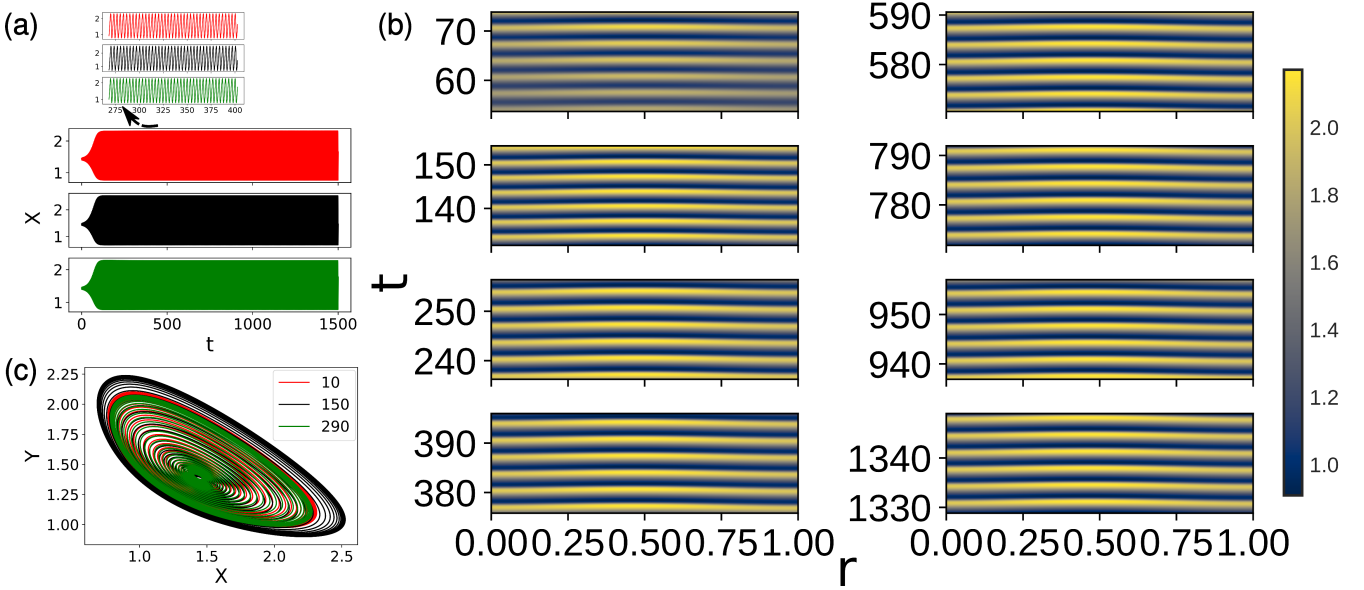}
			\caption{\label{fig14} For $\nu_0=2.75$ and $\nu_b=2.80$ and asymmetric influx (a)concentration profiles have homogeneous oscillation at all the three regimes in the presence of diffusion coefficients $D_{11}=D_{22}=0.0025$. (b)Spatiotemporal images have no evidence of wave direction change. (c)Phase portraits produce a limit cycle after the transient dynamics.}
		\end{center}
	\end{figure*}
	\begin{figure*}
		\begin{center}
			\includegraphics[width=\textwidth]{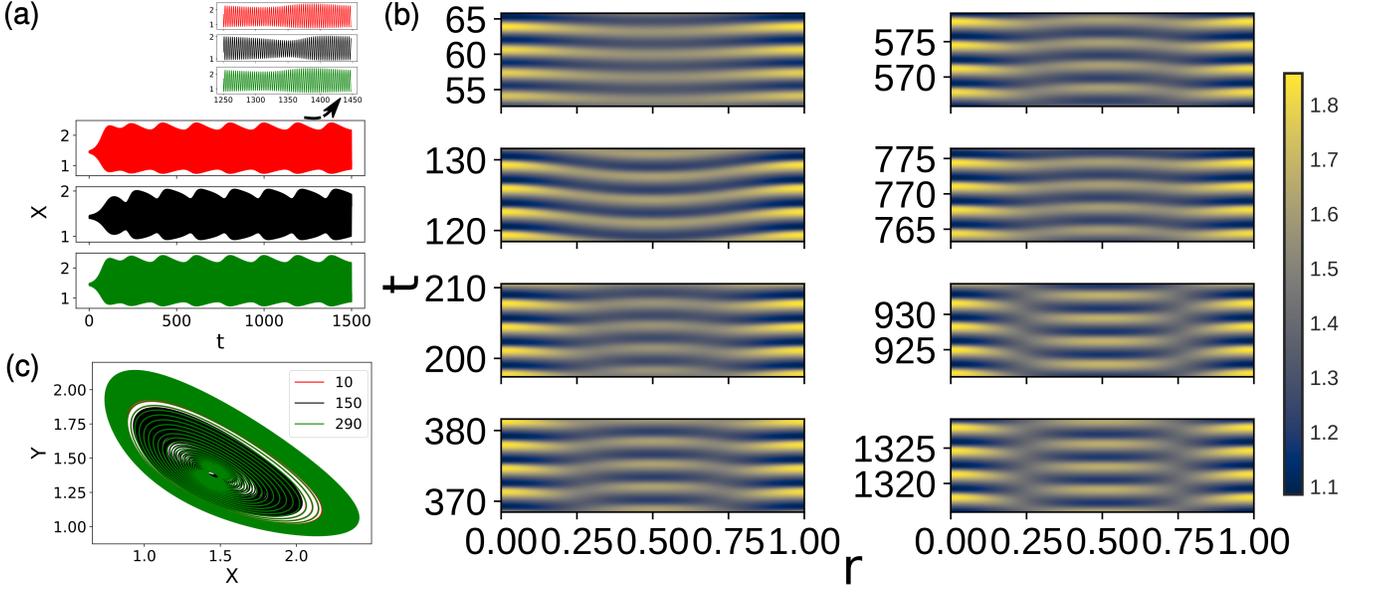}
			\caption{\label{fig15}  For symmetric parabolic influx profile with $\nu_0=2.80$ and $\nu_b=2.75$ and diffusion coefficients, $D_{11}=D_{22}=0.00051$ (a)we have amplitude modulated profiles in all three regimes. (b)Spatiotemporal images show that the oscillation at the center regime passes through the occasional flip. (c)The amplitude modulation nature is reflected in the dense trajectory of the phase portrait.}
		\end{center}
	\end{figure*}
	\begin{figure*}
		\begin{center}
			\includegraphics[width=\textwidth]{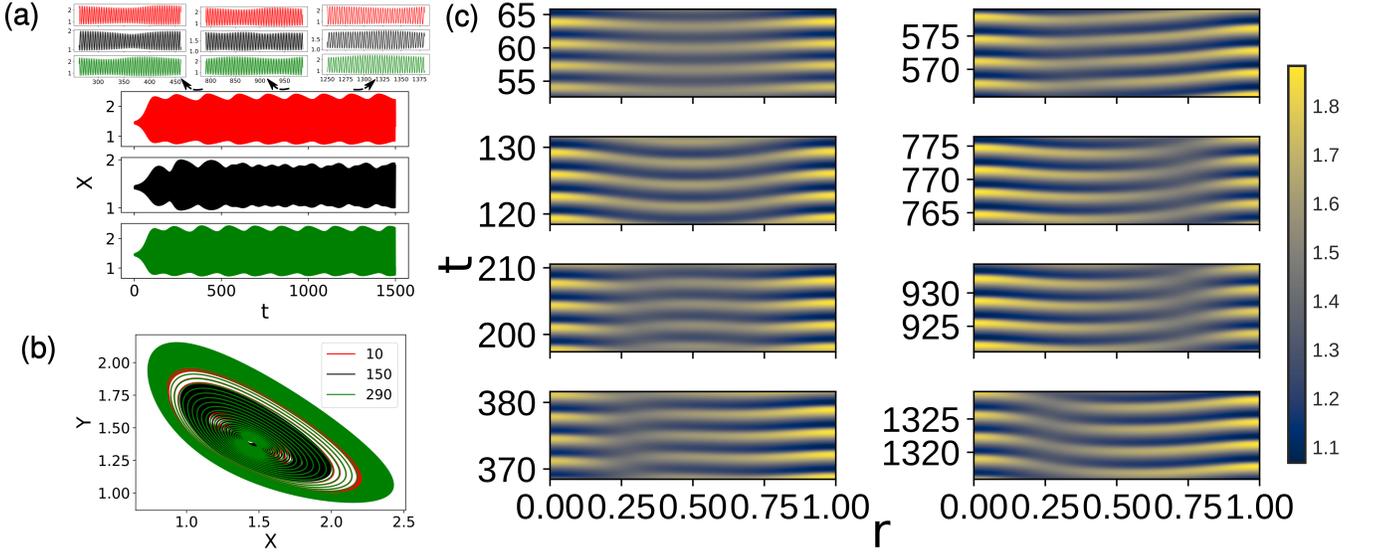}
			\caption{\label{fig16} For slightly asymmetric flux($r_0=0.495$) with $\nu_0=2.80$ and $\nu_b=2.75$ and diffusion coefficients, $D_{11}=D_{22}=0.00051$, (a)we have dynamic modulation strength at the center regime of the concentration profile.  (b)Spatiotemporal images demonstrate a wave direction switch. (c)The corresponding phase portrait is illustrated. }
		\end{center}
	\end{figure*}
	\begin{figure*}
		\begin{center}
			\includegraphics[width=\textwidth]{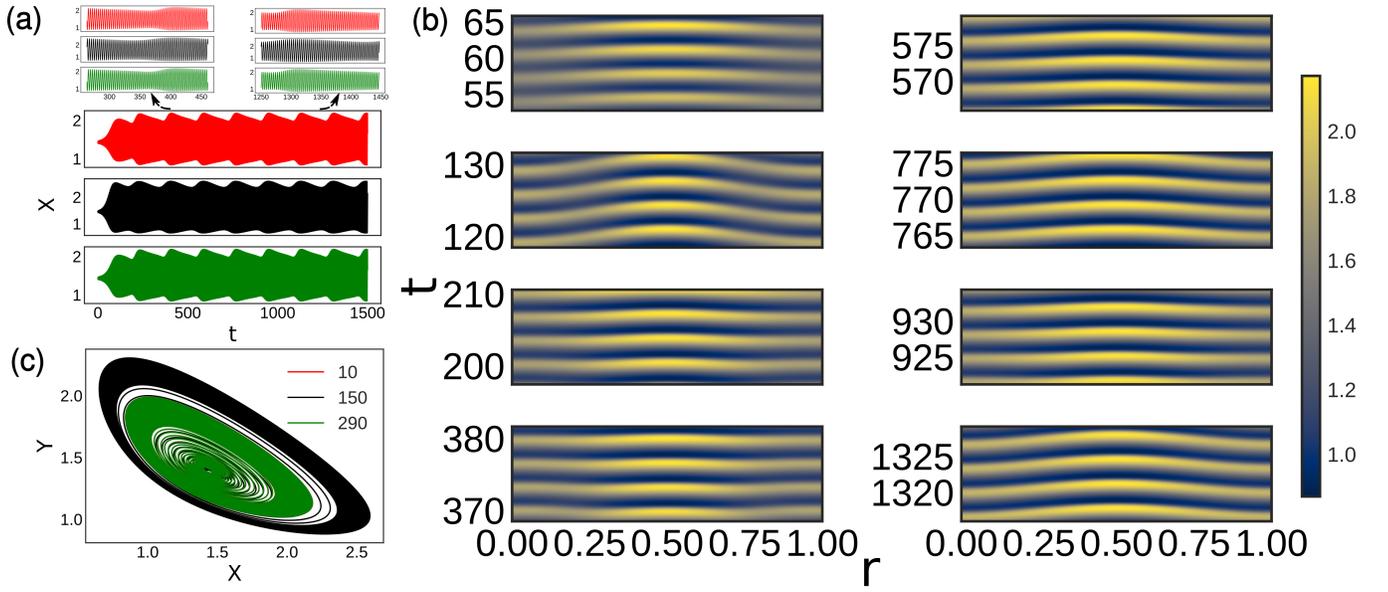}
			\caption{\label{fig17} Symmetric influx with $\nu_0=2.75$ and $\nu_b=2.80$, and diffusion coefficients, $D_{11}=D_{22}=0.00051$, (a)we have amplitude modulation profile in all three regimes.(b)No wave direction change in the spatiotemporal image. (c)Corresponding phase portraits are provided.}
		\end{center}
	\end{figure*}
	\begin{figure*}
		\begin{center}
			\includegraphics[width=\textwidth]{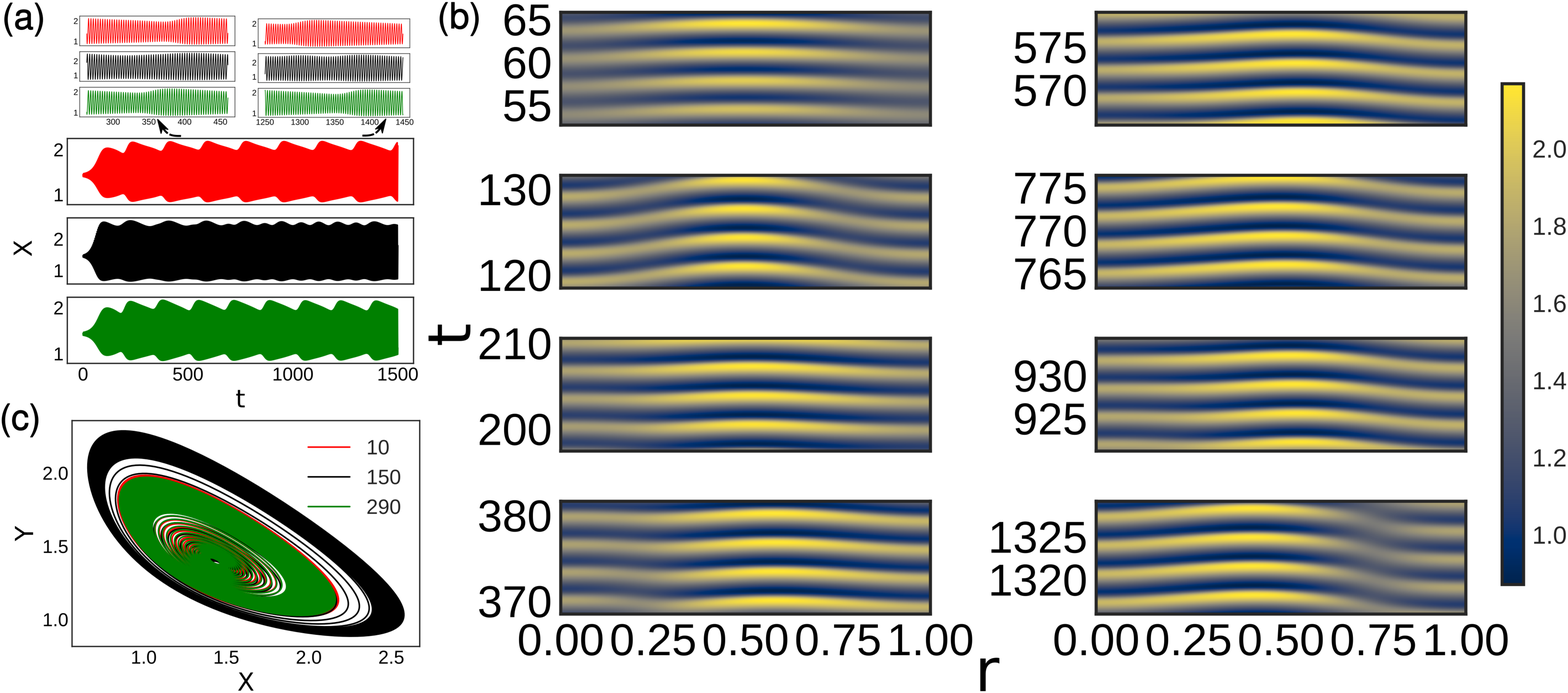}
			\caption{\label{fig18} For asymmetric flux with $\nu_0=2.75$ and $\nu_b=2.80$, and diffusion coefficients, $D_{11}=D_{22}=0.00051$, (a)time-varying amplitude modulation emerges at the center regime. (b)Spatiotemporal images demonstrate occasional wave propagation direction changes. (c)A corresponding phase portrait is shown.}
		\end{center}
	\end{figure*}
	
	In the Selkov model with the homogeneous influx,  we expect harmonic oscillation behavior for the control parameter value near the onset of Hopf instability and relaxation oscillation behavior for the control parameter away from the onset point of the instability. However, with the spatially inhomogenous distribution of the control parameter, the possibility of simultaneously having control parameter values related to completely different oscillation behaviors arises. So, we can have two different or the same oscillatory behaviors near the center and spatial domain boundaries for the inhomogeneous parabolic influx. Depending on the oscillation nature of the concentration dynamics near the center and boundaries, we have carried out our whole investigation for the following scenarios:    
	\begin{itemize}
		\item Center- relaxation oscillation and boundaries- relaxation oscillation
		\item Center- relaxation oscillation and boundaries- harmonic oscillation
		\item Center- harmonic oscillation and boundaries- relaxation oscillation
		\item Center- harmonic oscillation and boundaries: harmonic oscillation.
	\end{itemize}  
	These scenarios can be realized by specifying appropriate values of $\nu_0$ and $\nu_b$ in eq. \eqref{inhomoflux}. We would extensively look into these scenarios for two different orders of the diffusion coefficients, i.e., $D_{11}=D_{22}=0.0025$ and $D_{11}=D_{22}=0.00051$ to reveal the role played by the diffusion in dictating the pattern of the concentration dynamics. Further, for these scenarios and diffusion coefficients, we would implement two slightly different parabolic inflow of the chemostatted species with $r_{0}=0.5$ and $r_{0}=0.495$ in eq. \eqref{inhomoflux} for spatially symmetric and asymmetric parabolic influx profiles, respectively.  
	
	\textbf{Center- relaxation oscillation and boundaries- relaxation oscillation:} 
	
	For a homogeneous influx of $\nu$, the concentration dynamics of the system have relaxation oscillation for $\nu=2.55$ and $\nu=2.60$ in the presence of two equal diffusion coefficients of the species. This concentration behavior remains intact for two different order of the diffusion coefficients incorporated in this study, i.e., $D_{11}=D_{22}=0.0025$ and $D_{11}=D_{22}=0.00051$. So, we choose $\nu_0=2.55$ and $\nu_b=2.60$ in the parabolic inflow profile of $\nu$ to set relaxation oscillation behavior near the center and two boundaries of the spatial domain. 
	
	In fig. \ref{fig2}, time series of $X$ concentration within the range of $t=0$ to $t=1500$  and corresponding phase portrait of $X$ and $Y$ concentration dynamics has been illustrated for $D_{11}=D_{22}=0.0025$. We obtain three different profiles of the concentration dynamics in fig. \ref{fig2}(a) and (b)  by specifying the index of $r$ as $10$, $150$, and $290$, respectively. These three indices are associated with the different spatial oscillatory regions near the left, center, and right boundary, respectively. In this investigation, unless otherwise stated, we would use the same spatial points for all the time series and phase portraits. The time series and phase portrait in fig. \ref{fig2}(a) and (b) suggest that concentration dynamics have periodic oscillation for all three regimes. The nature of oscillation can be viewed in the subplot of fig. \ref{fig2}(a)(indicated by an arrow) depicted for a short time window.
	
	As we weaken the order of diffusion coefficients to $D_{11}=D_{22}=0.00051,$ keeping all other parameters fixed, complicated concentration dynamics arise near the center regime of the system, as shown in Fig. \ref{fig3}(a). A relatively small time window in the subplot captures an intricate mixed oscillatory mode of the concentration dynamics near the center regimes. This oscillatory pattern yields a quasiperiodic behavior for a long time range. A dense phase space trajectory beyond the transient spiral curve also indicates the quasiperiodic nature of dynamics. This nature of concentration dynamics at the center part of the spatial domain is reflected in the spatiotemporal images at the different time intervals in \ref{fig3} (b). However, the temporal concentration dynamics near the boundaries exhibit an almost periodic oscillation. In Fig \ref{fig2} and \ref{fig3}, we have considered a symmetric parabolic influx with $r_0=0.5$ and the phase portraits demonstrate that oscillations at both boundaries are completely synchronized for this symmetric parabolic influx. Comparing Fig. \ref{fig2} and \ref{fig3}, we assert that although the diffusion coefficients have a very low magnitude, they still play a crucial role in dictating the dynamics of the system. In particular, higher diffusion coefficients in Fig. \ref{fig2} suppresses the intricate temporal behavior near the spatial center of the system in Fig. ~\ref{fig3}.                              
	
	Now we introduce a slight asymmetry in the parabolic inflow of the chemostatted species by considering $r_{0}=0.495$ in eq. \eqref{inhomoflux} to encapsulate its effect on the concentration dynamics. Essentially, the asymmetric parabolic inflow hampers the synchrony between oscillations at two boundaries, as evident from the phase portrait in Fig.~\ref{fig4} and \ref{fig5}(c). Additionally, this slight asymmetry in the influx induces a dynamic transition from almost periodic oscillation(subplot on the left in Fig.~\ref{fig4}(a)) to the complex oscillation(subplot on the right) in concentration dynamics near the center of the spatial domain, even in the case of higher diffusivity($D_{11}=D_{22}=0.0025$). The complex oscillation mode brings quasiperiodic dynamics near the center regime of the concentration for a higher time range as supported by the dense phase space trajectories in Fig.~\ref{fig4}(c). So as an effect of asymmetric chemostatted species inflow, the suppression by diffusion dilutes over time, and we have a dynamic transition from periodic oscillation to quasiperiodic oscillation near the center of the system. From spatiotemporal images in Fig.~\ref{fig4}(b), we notice that waves initially propagate inwardly from the border (the first three images on the left panel). Then the propagation direction of waves gradually changes from right to left border and remains the same for the rest of the time. This wave direction change is introduced as an effect of the weak asymmetry in the inflow of the chemostatted species. Now for the lower diffusion coefficients, $D_{11}=D_{22}=0.00051$, the asymmetry in the parabolic influx induces some sudden drop in the periodic oscillation amplitude of the concentration dynamics near the left border in Fig.~\ref{fig5}(a). Moreover, the quasiperiodic oscillation of the symmetric influx near the center spatial regime loses its form, and a chaotic oscillation appears over the whole time range. Spatiotemporal images in Fig.~\ref{fig5} (b) suggest that although more complicated wave profiles appear owing to the chaotic oscillation at the center regime, the nature of wave propagation direction change remains the same as in the case of the higher order diffusion.       
	
	\textbf{Center-relaxation oscillation and boundaries-harmonic oscillation:}
	This scenario with relaxation oscillations near the center regimes and harmonic oscillations near two boundaries is more interesting than the previous one due to two completely different types of oscillations at the border and center regimes of the system. To realize this situation, we have set $\nu_0=2.55$ and $\nu_b=2.75$ in the parabolic influx equation. Now, like in the previous scenario, we would investigate the effect of the diffusion and the symmetry of the parabolic influx on the concentration dynamics in detail. 
	
	In higher diffusive cases with $D_{11}=D_{22}=0.0025$, the time series of the concentration in all three regimes exhibit the amplitude modulation  for both the symmetric($r_0=0.5$) and asymmetric($r_0=0.495$) parabolic inflow of the chemostatted species as evident from subplots of Fig.~\ref{fig6}(a) and Fig. ~\ref{fig7}(a), respectively. These amplitude modulations arise from the generation of two different frequency waves in the presence of harmonic- and relaxation-type oscillation. In both cases, modulation is higher for the border regimes relative to the center regimes. All the regimes have a quasiperiodic nature for the long time range, as suggested by the dense phase portrait in Fig.~\ref{fig6} and \ref{fig7}. In particular, the quasiperiodicity at the center regime is highly intricate for the asymmetric influx. The spatiotemporal images of the inwardly traveling waves now have occasional signatures of the coexistence of high and low frequency as we observe the different wave numbers near the border and center in Fig.~\ref{fig6}(b). However, there is no change in propagation direction for the symmetric influx case. The change in propagation direction is generated as we bring asymmetry within the inflow profile of the chemostatted species. Indeed, the wave propagation passes through several changes in direction over the whole time range in the presence of asymmetry. Within a very short time interval, the inwardly traveling wave profile changes its direction slightly as it moves from the left to right border and then switches from the right to the left border before resorting to the inwardly moving nature again. Despite the complex nature of the wave profile, we can spot this sequence of wave direction change several times. Some of these direction switches are captured by the spatiotemporal images in Fig. ~\ref{fig7}(b).                                                   
	
	For lower diffusion coefficients of $D_{11}=D_{22}=0.00051$ in Fig.~\ref{fig8}, with symmetric influx profiles, modulated oscillatory behavior is stronger at the center regime than in the boundary ones. Due to the relatively low diffusion, the complexity of the dynamics is more prominent here relative to the previous one. Despite no wave direction shift, multifold wave profiles emerge in the system due to the complex quasiperiodic oscillations at the center regimes, as can be noticed from the spatiotemporal images in Fig.~\ref{fig8}(b). In the presence of asymmetric parabolic influx of chemostatted species, the strong amplitude modulation emerges at the center, and the whole spatiotemporal profiles in Fig.~\ref{fig9}(b) become more intricate. More interestingly, the temporal dynamics of the concentration at the center regime demonstrate a quasiperiodic route to chaos~\cite{experimentquasiperiodic}. At the boundaries, we obtain highly complex quasiperiodic oscillation. Further, we notice a slight shift in the wave direction in some time intervals in Fig.~\ref{fig9}(b). However, the shift in the wave direction is not as apparent compared to the case of the higher diffusion coefficients of $D_{11}=D_{22}=0.0025$ in this scenario. This difference arises due to the dominance of highly complicated chaotic flow near the center in the case of lower diffusion coefficients.                 
	
	\textbf{Center-harmonic oscillation and boundaries-relaxation oscillation:}
	We now would investigate the opposite scenario of the previous one, i.e., the harmonic oscillation at the center and the relaxation oscillator at the boundaries. For this situation, we consider $\nu_0=2.75$ and $\nu_b=2.55$ in the parabolic inflow equation and thus principally flip the parabolic influx profile of the chemostatted species. 
	
	For the higher diffusion coefficients of $D_{11}=D_{22}=0.0025$ with symmetric inflow, the amplitude modulation at the center of the spatial domain portrays a beat interference pattern in the concentration profiles as apparent from the subplot of Fig.~\ref{fig10}. This type of beat phenomenon has also been observed in the chemical system of periodically forced pH oscillator~\cite{beatphoscillator} or globally coupled continuum chemical oscillator system~\cite{pkgg3}. The dynamics of this state display the torus phase space corresponding to the quasiperiodic flow near its death. Oscillations near the border have small amplitude modulation with long-time dynamics lying on a toroidal surface. The spatiotemporal images of concentration demonstrate outwardly moving traveling waves without any dynamic direction change of the wave over time in Fig.~\ref{fig10}(b). Whereas, in the asymmetric influx case in Fig.~\ref{fig11}, there are three different appearances of time series on three different time ranges. Initially(subplot on the left), we have amplitude modulation in all three spatial regimes. Then as time increases(subplot in the middle), a beat pattern develops in the center regimes of the system, and amplitude modulation fades away in the oscillations near two boundaries of the spatial domain. Then almost periodic oscillatory behavior emerges in all three regimes for a higher temporal range(subplot on the right). Over the time range of interest, these different behaviors bear the signature of dynamic quasiperiodic to periodic transition. Although slightly different wave profiles are generated near the center of the spatial regime, spatiotemporal images in Fig.~\ref{fig11}(b) suggest that only outwardly moving waves are present in the system.                                 
	
	As we decrease diffusion coefficients to $D_{11}=D_{22}=0.00051$ and employ a symmetric parabolic influx of chemostatted species, a periodic oscillatory behavior near the center and quasiperiodic oscillation near two boundaries appear in the time series of the concentration. Despite some spatiotemporal images exhibiting complex wave profiles, waves always move outwardly from the center regimes in Fig.~\ref{fig12}. Implementing an asymmetric parabolic influx does not change the temporal behavior of any spatial regimes as the short time range oscillatory profile shown in the subplot of Fig.~\ref{fig13}(a) suggests. However, the close examination of the spatiotemporal profiles in Fig.~\ref{fig13}(b) reveals the change in the wave profile propagating direction(look at the difference between the time interval $370-380$ and $765-775$). This result with asymmetric parabolic influx is contrary to the case of higher diffusion coefficients and thus again discloses the crucial role of diffusion.                  
	
	\textbf{Center-harmonic oscillation and boundaries-harmonic oscillation:}
	In the last scenario, we define our parabolic influx to obtain harmonic oscillators near the boundaries and the center regimes. Although the oscillatory natures of the two boundaries and center regimes are the same, it is evident from Fig. \ref{fig1} (a) and (d) that we can have completely different phase profile variations concerning time depending on whether the parabolic profile of the influx opens upward or downward. Particularly, without considering the diffusion, if we set $\nu_0=2.80$ and $\nu_b=2.75$ in the parabolic influx equation, Fig. \ref{fig1} (a) suggests phase profile gets reversed with time. Whereas, for $\nu_0=2.75$ and $\nu_b=2.80$ in the parabolic inflow of the chemostatted species, the phase in Fig. \ref{fig1} (d) illustrates no phase reversal. So we are going to explore these two sets of values.   
	
	For $\nu_0=2.80$ and $\nu_b=2.75$ and  higher diffusion coefficients $D_{11}=D_{22}=0.0025$, all three regimes present homogeneous oscillation(not shown here). We also do not acquire any wave direction change by the asymmetry in the influx profile. Due to asymmetry, only the concentration profiles at two boundaries are less synchronized than in the symmetric case. Now for $\nu_0=2.75$ and $\nu_b=2.80$ with the same diffusion coefficients, we again have similar homogenous oscillation and limit cycle in the phase space after the transient response in Fig.~\ref{fig14}. Also, no wave direction change is noted in the symmetric and asymmetric influx profiles(see spatiotemporal images in Fig. \ref{fig14} (b)). So it is evident that two different variations of the phase profile(without diffusion) have no effect on changing the wave direction or other qualitative wave behavior in the presence of these diffusion coefficients.                         
	
	Whereas employing a lower order of the diffusion coefficient,  $D_{11}=D_{22}=0.00051$, we acquire amplitude modulation profiles for all three regimes with $\nu_0=2.80$ and $\nu_b=2.75$ in symmetric parabolic influx case in Fig.~\ref{fig15}. The amplitude modulation nature is also reflected in the dense phase space trajectory for all the regimes. The quasiperiodic nature of all three regimes is very complex. More importantly, the oscillation at the center spatial part of the system passes through the occasional flip of the parabolic profile. Now, with the asymmetry in the parabolic influx, we have a wave direction switch contrary to the higher diffusion coefficient order, as seen from the spatiotemporal illustrations in Fig.~\ref{fig16} (b). The temporal dynamics of three different regimes also suggest that all three regimes have an amplitude-modulated wave profile, with the center regime having dynamic modulation strengths.                                 
	
	With $\nu_0=2.75$ and $\nu_b=2.80$ in the symmetric parabolic inflow of the chemostatted species and the same lower diffusion coefficients, amplitude modulated wave pattern in all three regimes emerge in Fig. \ref{fig17}. For the asymmetric case, we again obtain a time-varying amplitude modulation near the central regime of the system in Fig. \ref{fig18} (a). The asymmetry also brings occasional wave propagation direction changes in the system, as evident from the spatiotemporal images in Fig. \ref{fig18} (b).
	
	\section{\label{Con}Conclusions}
Through a systematic investigation {of a prototypical biological oscillator, we have identified the possible spatial coexistence of different temporal behaviors related to the allocation of the control parameter values over the spatial domain. These coexisting nontrivial temporal behaviors, in the presence of the inhomogeneous influx, depend on the nature of oscillations near the center and boundary regimes of the system and interaction between corresponding waves. Numerical results concerning four possible scenarios are presented to illuminate this connection.}  Further, our investigation reveals that the symmetry property of the influx of the chemostatted species mainly decides the fate of the direction of glycolytic wave propagation. {More specifically, we detect that the direction change of the traveling wave is feasible only with slight asymmetry in the inhomogeneous influx. Moreover,} the crucial role of diffusion in dictating the system dynamics is illustrated by taking two different orders of diffusion coefficients. {For two dissimilar diffusion coefficient orders, we demonstrate that different spatial dynamics can emerge for the same influx distribution. Thus we can have distinct natures in wave propagation direction switches for the same inhomogeneous influx. The results of this study would aid us in engineering the dynamics of the system according to a specific purpose.}  The rich variety of states comprising the periodic, quasiperiodic, and chaotic dynamics demonstrated here in different spatial regimes of the system can also be confirmed by the standard Poincare maps method~\cite{hirsch2012differential}. Such dynamical variety in the Selkov model is recently reported by incorporating periodic influx~\cite{postoperiodic} in the absence of diffusion. {These results can be associated with the experimental results for the dynamics of glycolytic waves in an open spatial reactor.}   
	
Besides enriching the previous studies of traveling wave propagation in a system with spatially heterogeneous parameter~\cite{LAVROVA2009127, Lavrova2009PhaseInflux, postoperiodic, PAGE200595}, this complete theoretical depiction lays a proper dynamical basis for investigating the thermodynamic evolution of such a system in the presence of nonuniform parameters. Beyond the glycolysis, this study is equally relevant for gaining an insight into other large variety of biochemical oscillators of generalized Rayleigh oscillator~\cite{SGDSR} class in the presence of inhomogeneous control parameters. Moreover, the coexistence of various nonlinear dynamical states and temporal transition between different states found here in the presence of inhomogeneous influx bear a resemblance to the characteristics of dynamical diseases~\cite{dynamicdise1, dynamicdise2, dynamicdise3} and thus can be an important aspect to explore. As the cellular spatiotemporal oscillation behaviors rely on diffusion and the cell's location in the array~\cite{SCHUTZE2010104}, our findings can have crucial implications in the context of biological information processing and spreading~\cite{PETTY2006217, PURVIS2013945, BEHAR2010684}.  
\section*{Conflict of Interest}
The authors declare no conflict of interest.

\vspace{8mm}	
	\noindent\textsf{\textbf{Keywords}: Chaos, Glycolytic waves, Quasiperiodicity, Reaction-diffusion model, Selkov model} 
	\setlength{\bibsep}{0.0cm}   	
	\bibliographystyle{wiley-chemistry}       
	\bibliography{selkovreport}

\begin{thebibliography}{26}%
\makeatletter
\providecommand \@ifxundefined [1]{%
 \@ifx{#1\undefined}
}%
\providecommand \@ifnum [1]{%
 \ifnum #1\expandafter \@firstoftwo
 \else \expandafter \@secondoftwo
 \fi
}%
\providecommand \@ifx [1]{%
 \ifx #1\expandafter \@firstoftwo
 \else \expandafter \@secondoftwo
 \fi
}%
\providecommand \natexlab [1]{#1}%
\providecommand \enquote  [1]{``#1''}%
\providecommand \bibnamefont  [1]{#1}%
\providecommand \bibfnamefont [1]{#1}%
\providecommand \citenamefont [1]{#1}%
\providecommand \href@noop [0]{\@secondoftwo}%
\providecommand \href [0]{\begingroup \@sanitize@url \@href}%
\providecommand \@href[1]{\@@startlink{#1}\@@href}%
\providecommand \@@href[1]{\endgroup#1\@@endlink}%
\providecommand \@sanitize@url [0]{\catcode `\\12\catcode `\$12\catcode
  `\&12\catcode `\#12\catcode `\^12\catcode `\_12\catcode `\%12\relax}%
\providecommand \@@startlink[1]{}%
\providecommand \@@endlink[0]{}%
\providecommand \url  [0]{\begingroup\@sanitize@url \@url }%
\providecommand \@url [1]{\endgroup\@href {#1}{\urlprefix }}%
\providecommand \urlprefix  [0]{URL }%
\providecommand \Eprint [0]{\href }%
\providecommand \doibase [0]{https://doi.org/}%
\providecommand \selectlanguage [0]{\@gobble}%
\providecommand \bibinfo  [0]{\@secondoftwo}%
\providecommand \bibfield  [0]{\@secondoftwo}%
\providecommand \translation [1]{[#1]}%
\providecommand \BibitemOpen [0]{}%
\providecommand \bibitemStop [0]{}%
\providecommand \bibitemNoStop [0]{.\EOS\space}%
\providecommand \EOS [0]{\spacefactor3000\relax}%
\providecommand \BibitemShut  [1]{\csname bibitem#1\endcsname}%
\let\auto@bib@innerbib\@empty
%</preamble>
\bibitem [{\citenamefont {Falasco}\ \emph {et~al.}(2018)\citenamefont
  {Falasco}, \citenamefont {Rao},\ and\ \citenamefont
  {Esposito}}]{Falasco2018InformationPatterns}%
  \BibitemOpen
  \bibfield  {author} {\bibinfo {author} {\bibfnamefont {G.}~\bibnamefont
  {Falasco}}, \bibinfo {author} {\bibfnamefont {R.}~\bibnamefont {Rao}},\ and\
  \bibinfo {author} {\bibfnamefont {M.}~\bibnamefont {Esposito}},\ }\bibfield
  {title} {\bibinfo {title} {{Information Thermodynamics of Turing Patterns}},\
  }\href {https://doi.org/10.1103/PhysRevLett.121.108301} {\bibfield  {journal}
  {\bibinfo  {journal} {Physical Review Letters}\ }\textbf {\bibinfo {volume}
  {121}},\ \bibinfo {pages} {108301} (\bibinfo {year} {2018})}\BibitemShut
  {NoStop}%
\bibitem [{\citenamefont {Avanzini}\ \emph {et~al.}(2019)\citenamefont
  {Avanzini}, \citenamefont {Falasco},\ and\ \citenamefont
  {Esposito}}]{thermodynamicschemifcalwaves}%
  \BibitemOpen
  \bibfield  {author} {\bibinfo {author} {\bibfnamefont {F.}~\bibnamefont
  {Avanzini}}, \bibinfo {author} {\bibfnamefont {G.}~\bibnamefont {Falasco}},\
  and\ \bibinfo {author} {\bibfnamefont {M.}~\bibnamefont {Esposito}},\
  }\bibfield  {title} {\bibinfo {title} {Thermodynamics of chemical waves},\
  }\href {https://doi.org/10.1063/1.5126528} {\bibfield  {journal} {\bibinfo
  {journal} {The Journal of Chemical Physics}\ }\textbf {\bibinfo {volume}
  {151}},\ \bibinfo {pages} {234103} (\bibinfo {year} {2019})},\ \Eprint
  {https://arxiv.org/abs/https://doi.org/10.1063/1.5126528}
  {https://doi.org/10.1063/1.5126528} \BibitemShut {NoStop}%
\bibitem [{\citenamefont {Kumar}\ and\ \citenamefont
  {Gangopadhyay}(2020)}]{pkgg}%
  \BibitemOpen
  \bibfield  {author} {\bibinfo {author} {\bibfnamefont {P.}~\bibnamefont
  {Kumar}}\ and\ \bibinfo {author} {\bibfnamefont {G.}~\bibnamefont
  {Gangopadhyay}},\ }\bibfield  {title} {\bibinfo {title} {Energetic and
  entropic cost due to overlapping of turing-hopf instabilities in the presence
  of cross diffusion},\ }\href {https://doi.org/10.1103/PhysRevE.101.042204}
  {\bibfield  {journal} {\bibinfo  {journal} {Phys. Rev. E}\ }\textbf {\bibinfo
  {volume} {101}},\ \bibinfo {pages} {042204} (\bibinfo {year}
  {2020})}\BibitemShut {NoStop}%
\bibitem [{\citenamefont {Kumar}\ and\ \citenamefont
  {Gangopadhyay}(2021)}]{pkgg2}%
  \BibitemOpen
  \bibfield  {author} {\bibinfo {author} {\bibfnamefont {P.}~\bibnamefont
  {Kumar}}\ and\ \bibinfo {author} {\bibfnamefont {G.}~\bibnamefont
  {Gangopadhyay}},\ }\bibfield  {title} {\bibinfo {title} {Nonequilibrium
  thermodynamics of glycolytic traveling wave: Benjamin-feir instability},\
  }\href {https://doi.org/10.1103/PhysRevE.104.014221} {\bibfield  {journal}
  {\bibinfo  {journal} {Phys. Rev. E}\ }\textbf {\bibinfo {volume} {104}},\
  \bibinfo {pages} {014221} (\bibinfo {year} {2021})}\BibitemShut {NoStop}%
\bibitem [{\citenamefont {Kumar}\ and\ \citenamefont
  {Gangopadhyay}(2022)}]{pkgg3}%
  \BibitemOpen
  \bibfield  {author} {\bibinfo {author} {\bibfnamefont {P.}~\bibnamefont
  {Kumar}}\ and\ \bibinfo {author} {\bibfnamefont {G.}~\bibnamefont
  {Gangopadhyay}},\ }\bibfield  {title} {\bibinfo {title} {Nonequilibrium
  thermodynamic characterization of chimeras in a continuum chemical oscillator
  system},\ }\href {https://doi.org/10.1103/PhysRevE.105.034208} {\bibfield
  {journal} {\bibinfo  {journal} {Phys. Rev. E}\ }\textbf {\bibinfo {volume}
  {105}},\ \bibinfo {pages} {034208} (\bibinfo {year} {2022})}\BibitemShut
  {NoStop}%
\bibitem [{\citenamefont {Argoul}\ \emph {et~al.}(1987)\citenamefont {Argoul},
  \citenamefont {Arneodo}, \citenamefont {Richetti},\ and\ \citenamefont
  {Roux}}]{experimentquasiperiodic}%
  \BibitemOpen
  \bibfield  {author} {\bibinfo {author} {\bibfnamefont {F.}~\bibnamefont
  {Argoul}}, \bibinfo {author} {\bibfnamefont {A.}~\bibnamefont {Arneodo}},
  \bibinfo {author} {\bibfnamefont {P.}~\bibnamefont {Richetti}},\ and\
  \bibinfo {author} {\bibfnamefont {J.~C.}\ \bibnamefont {Roux}},\ }\bibfield
  {title} {\bibinfo {title} {From quasiperiodicity to chaos in the
  belousov–zhabotinskii reaction. i. experiment},\ }\href
  {https://doi.org/10.1063/1.452751} {\bibfield  {journal} {\bibinfo  {journal}
  {The Journal of Chemical Physics}\ }\textbf {\bibinfo {volume} {86}},\
  \bibinfo {pages} {3325} (\bibinfo {year} {1987})},\ \Eprint
  {https://arxiv.org/abs/https://doi.org/10.1063/1.452751}
  {https://doi.org/10.1063/1.452751} \BibitemShut {NoStop}%
\bibitem [{\citenamefont {SEL'KOV}(1968)}]{selkov}%
  \BibitemOpen
  \bibfield  {author} {\bibinfo {author} {\bibfnamefont {E.~E.}\ \bibnamefont
  {SEL'KOV}},\ }\bibfield  {title} {\bibinfo {title} {Self-oscillations in
  glycolysis 1. a simple kinetic model},\ }\href
  {https://doi.org/https://doi.org/10.1111/j.1432-1033.1968.tb00175.x}
  {\bibfield  {journal} {\bibinfo  {journal} {European Journal of
  Biochemistry}\ }\textbf {\bibinfo {volume} {4}},\ \bibinfo {pages} {79}
  (\bibinfo {year} {1968})}\BibitemShut {NoStop}%
\bibitem [{\citenamefont {Richter}\ \emph {et~al.}(1981)\citenamefont
  {Richter}, \citenamefont {Rehmus},\ and\ \citenamefont
  {Ross}}]{reversibleselkoov}%
  \BibitemOpen
  \bibfield  {author} {\bibinfo {author} {\bibfnamefont {P.~H.}\ \bibnamefont
  {Richter}}, \bibinfo {author} {\bibfnamefont {P.}~\bibnamefont {Rehmus}},\
  and\ \bibinfo {author} {\bibfnamefont {J.}~\bibnamefont {Ross}},\ }\bibfield
  {title} {\bibinfo {title} {{Control and Dissipation in Oscillatory Chemical
  Engines}},\ }\href {https://doi.org/10.1143/PTP.66.385} {\bibfield  {journal}
  {\bibinfo  {journal} {Progress of Theoretical Physics}\ }\textbf {\bibinfo
  {volume} {66}},\ \bibinfo {pages} {385} (\bibinfo {year} {1981})}\BibitemShut
  {NoStop}%
\bibitem [{\citenamefont {Strogatz}(2018)}]{strogatznonlinear}%
  \BibitemOpen
  \bibfield  {author} {\bibinfo {author} {\bibfnamefont {S.~H.}\ \bibnamefont
  {Strogatz}},\ }\href@noop {} {\emph {\bibinfo {title} {Nonlinear dynamics and
  chaos: with applications to physics, biology, chemistry, and engineering}}}\
  (\bibinfo  {publisher} {CRC press},\ \bibinfo {year} {2018})\BibitemShut
  {NoStop}%
\bibitem [{\citenamefont {Aranson}\ and\ \citenamefont
  {Kramer}(2002)}]{aranson2002world}%
  \BibitemOpen
  \bibfield  {author} {\bibinfo {author} {\bibfnamefont {I.~S.}\ \bibnamefont
  {Aranson}}\ and\ \bibinfo {author} {\bibfnamefont {L.}~\bibnamefont
  {Kramer}},\ }\bibfield  {title} {\bibinfo {title} {The world of the complex
  ginzburg-landau equation},\ }\href {https://doi.org/10.1103/RevModPhys.74.99}
  {\bibfield  {journal} {\bibinfo  {journal} {Reviews of Modern Physics}\
  }\textbf {\bibinfo {volume} {74}},\ \bibinfo {pages} {99} (\bibinfo {year}
  {2002})}\BibitemShut {NoStop}%
\bibitem [{\citenamefont {Cross}\ and\ \citenamefont
  {Greenside}(2009)}]{Cross2009PatternSystems}%
  \BibitemOpen
  \bibfield  {author} {\bibinfo {author} {\bibfnamefont {M.}~\bibnamefont
  {Cross}}\ and\ \bibinfo {author} {\bibfnamefont {H.}~\bibnamefont
  {Greenside}},\ }\href {https://doi.org/10.1017/CBO9780511627200} {\emph
  {\bibinfo {title} {{Pattern Formation and Dynamics in Nonequilibrium
  Systems}}}}\ (\bibinfo  {publisher} {Cambridge University Press},\ \bibinfo
  {address} {Cambridge},\ \bibinfo {year} {2009})\BibitemShut {NoStop}%
\bibitem [{\citenamefont {Nicolis}(1995)}]{Nicolis1995IntroductionScience}%
  \BibitemOpen
  \bibfield  {author} {\bibinfo {author} {\bibfnamefont {G.}~\bibnamefont
  {Nicolis}},\ }\href@noop {} {\emph {\bibinfo {title} {{Introduction to
  nonlinear science}}}}\ (\bibinfo  {publisher} {Cambridge University Press},\
  \bibinfo {year} {1995})\ p.\ \bibinfo {pages} {254}\BibitemShut {NoStop}%
\bibitem [{\citenamefont {Krylov}\ and\ \citenamefont
  {Bogoliubov}(1949)}]{krylov1949introduction}%
  \BibitemOpen
  \bibfield  {author} {\bibinfo {author} {\bibfnamefont {N.~M.}\ \bibnamefont
  {Krylov}}\ and\ \bibinfo {author} {\bibfnamefont {N.~N.}\ \bibnamefont
  {Bogoliubov}},\ }\href@noop {} {\emph {\bibinfo {title} {Introduction to
  non-linear mechanics}}}\ (\bibinfo  {publisher} {Princeton University
  Press},\ \bibinfo {year} {1949})\BibitemShut {NoStop}%
\bibitem [{\citenamefont {Lavrova}\ \emph
  {et~al.}(2009{\natexlab{a}})\citenamefont {Lavrova}, \citenamefont
  {Schimansky-Geier},\ and\ \citenamefont
  {Postnikov}}]{Lavrova2009PhaseInflux}%
  \BibitemOpen
  \bibfield  {author} {\bibinfo {author} {\bibfnamefont {A.~I.}\ \bibnamefont
  {Lavrova}}, \bibinfo {author} {\bibfnamefont {L.}~\bibnamefont
  {Schimansky-Geier}},\ and\ \bibinfo {author} {\bibfnamefont {E.~B.}\
  \bibnamefont {Postnikov}},\ }\bibfield  {title} {\bibinfo {title} {{Phase
  reversal in the Selkov model with inhomogeneous influx}},\ }\href
  {https://doi.org/10.1103/PhysRevE.79.057102} {\bibfield  {journal} {\bibinfo
  {journal} {Physical Review E - Statistical, Nonlinear, and Soft Matter
  Physics}\ }\textbf {\bibinfo {volume} {79}},\ \bibinfo {pages} {1} (\bibinfo
  {year} {2009}{\natexlab{a}})}\BibitemShut {NoStop}%
\bibitem [{\citenamefont {Benjamin}(1967)}]{benjamin1967instability}%
  \BibitemOpen
  \bibfield  {author} {\bibinfo {author} {\bibfnamefont {T.~B.}\ \bibnamefont
  {Benjamin}},\ }\bibfield  {title} {\bibinfo {title} {Instability of periodic
  wavetrains in nonlinear dispersive systems},\ }\href@noop {} {\bibfield
  {journal} {\bibinfo  {journal} {Proceedings of the Royal Society of London.
  Series A. Mathematical and Physical Sciences}\ }\textbf {\bibinfo {volume}
  {299}},\ \bibinfo {pages} {59} (\bibinfo {year} {1967})}\BibitemShut
  {NoStop}%
\bibitem [{\citenamefont {Kopell}\ and\ \citenamefont
  {Howard}(1973)}]{lamomega1}%
  \BibitemOpen
  \bibfield  {author} {\bibinfo {author} {\bibfnamefont {N.}~\bibnamefont
  {Kopell}}\ and\ \bibinfo {author} {\bibfnamefont {L.~N.}\ \bibnamefont
  {Howard}},\ }\bibfield  {title} {\bibinfo {title} {Plane wave solutions to
  reaction-diffusion equations},\ }\href
  {https://doi.org/https://doi.org/10.1002/sapm1973524291} {\bibfield
  {journal} {\bibinfo  {journal} {Studies in Applied Mathematics}\ }\textbf
  {\bibinfo {volume} {52}},\ \bibinfo {pages} {291} (\bibinfo {year}
  {1973})}\BibitemShut {NoStop}%
\bibitem [{\citenamefont {Lavrova}\ \emph
  {et~al.}(2009{\natexlab{b}})\citenamefont {Lavrova}, \citenamefont {Bagyan},
  \citenamefont {Mair}, \citenamefont {Hauser},\ and\ \citenamefont
  {Schimansky-Geier}}]{LAVROVA2009127}%
  \BibitemOpen
  \bibfield  {author} {\bibinfo {author} {\bibfnamefont {A.}~\bibnamefont
  {Lavrova}}, \bibinfo {author} {\bibfnamefont {S.}~\bibnamefont {Bagyan}},
  \bibinfo {author} {\bibfnamefont {T.}~\bibnamefont {Mair}}, \bibinfo {author}
  {\bibfnamefont {M.}~\bibnamefont {Hauser}},\ and\ \bibinfo {author}
  {\bibfnamefont {L.}~\bibnamefont {Schimansky-Geier}},\ }\bibfield  {title}
  {\bibinfo {title} {Modeling of glycolytic wave propagation in an open spatial
  reactor with inhomogeneous substrate influx},\ }\href
  {https://doi.org/https://doi.org/10.1016/j.biosystems.2009.04.005} {\bibfield
   {journal} {\bibinfo  {journal} {Biosystems}\ }\textbf {\bibinfo {volume}
  {97}},\ \bibinfo {pages} {127 } (\bibinfo {year}
  {2009}{\natexlab{b}})}\BibitemShut {NoStop}%
\bibitem [{\citenamefont {Petzold}(1983)}]{LSODA}%
  \BibitemOpen
  \bibfield  {author} {\bibinfo {author} {\bibfnamefont {L.}~\bibnamefont
  {Petzold}},\ }\bibfield  {title} {\bibinfo {title} {Automatic selection of
  methods for solving stiff and nonstiff systems of ordinary differential
  equations},\ }\href {https://doi.org/10.1137/0904010} {\bibfield  {journal}
  {\bibinfo  {journal} {SIAM J. Sci. Stat. Comput.}\ }\textbf {\bibinfo
  {volume} {4}},\ \bibinfo {pages} {136–148} (\bibinfo {year}
  {1983})}\BibitemShut {NoStop}%
\bibitem [{\citenamefont {Lawson}\ \emph {et~al.}(2020)\citenamefont {Lawson},
  \citenamefont {Holló}, \citenamefont {Horvath}, \citenamefont {Kitahata},\
  and\ \citenamefont {Lagzi}}]{beatphoscillator}%
  \BibitemOpen
  \bibfield  {author} {\bibinfo {author} {\bibfnamefont {H.~S.}\ \bibnamefont
  {Lawson}}, \bibinfo {author} {\bibfnamefont {G.}~\bibnamefont {Holló}},
  \bibinfo {author} {\bibfnamefont {R.}~\bibnamefont {Horvath}}, \bibinfo
  {author} {\bibfnamefont {H.}~\bibnamefont {Kitahata}},\ and\ \bibinfo
  {author} {\bibfnamefont {I.}~\bibnamefont {Lagzi}},\ }\bibfield  {title}
  {\bibinfo {title} {Chemical resonance, beats, and frequency locking in forced
  chemical oscillatory systems},\ }\href
  {https://doi.org/10.1021/acs.jpclett.0c00586} {\bibfield  {journal} {\bibinfo
   {journal} {The Journal of Physical Chemistry Letters}\ }\textbf {\bibinfo
  {volume} {11}},\ \bibinfo {pages} {3014} (\bibinfo {year} {2020})},\ \bibinfo
  {note} {pMID: 32216274},\ \Eprint
  {https://arxiv.org/abs/https://doi.org/10.1021/acs.jpclett.0c00586}
  {https://doi.org/10.1021/acs.jpclett.0c00586} \BibitemShut {NoStop}%
\bibitem [{\citenamefont {Hirsch}\ \emph {et~al.}(2012)\citenamefont {Hirsch},
  \citenamefont {Smale},\ and\ \citenamefont
  {Devaney}}]{hirsch2012differential}%
  \BibitemOpen
  \bibfield  {author} {\bibinfo {author} {\bibfnamefont {M.~W.}\ \bibnamefont
  {Hirsch}}, \bibinfo {author} {\bibfnamefont {S.}~\bibnamefont {Smale}},\ and\
  \bibinfo {author} {\bibfnamefont {R.~L.}\ \bibnamefont {Devaney}},\
  }\href@noop {} {\emph {\bibinfo {title} {Differential equations, dynamical
  systems, and an introduction to chaos}}}\ (\bibinfo  {publisher} {Academic
  press},\ \bibinfo {year} {2012})\BibitemShut {NoStop}%
\bibitem [{\citenamefont {Verveyko}\ \emph {et~al.}(2017)\citenamefont
  {Verveyko}, \citenamefont {Verisokin},\ and\ \citenamefont
  {Postnikov}}]{postoperiodic}%
  \BibitemOpen
  \bibfield  {author} {\bibinfo {author} {\bibfnamefont {D.~V.}\ \bibnamefont
  {Verveyko}}, \bibinfo {author} {\bibfnamefont {A.~Y.}\ \bibnamefont
  {Verisokin}},\ and\ \bibinfo {author} {\bibfnamefont {E.~B.}\ \bibnamefont
  {Postnikov}},\ }\bibfield  {title} {\bibinfo {title} {Mathematical model of
  chaotic oscillations and oscillatory entrainment in glycolysis originated
  from periodic substrate supply},\ }\href {https://doi.org/10.1063/1.4996554}
  {\bibfield  {journal} {\bibinfo  {journal} {Chaos: An Interdisciplinary
  Journal of Nonlinear Science}\ }\textbf {\bibinfo {volume} {27}},\ \bibinfo
  {pages} {083104} (\bibinfo {year} {2017})},\ \Eprint
  {https://arxiv.org/abs/https://doi.org/10.1063/1.4996554}
  {https://doi.org/10.1063/1.4996554} \BibitemShut {NoStop}%
\bibitem [{\citenamefont {Page}\ \emph {et~al.}(2005)\citenamefont {Page},
  \citenamefont {Maini},\ and\ \citenamefont {Monk}}]{PAGE200595}%
  \BibitemOpen
  \bibfield  {author} {\bibinfo {author} {\bibfnamefont {K.~M.}\ \bibnamefont
  {Page}}, \bibinfo {author} {\bibfnamefont {P.~K.}\ \bibnamefont {Maini}},\
  and\ \bibinfo {author} {\bibfnamefont {N.~A.}\ \bibnamefont {Monk}},\
  }\bibfield  {title} {\bibinfo {title} {Complex pattern formation in
  reaction–diffusion systems with spatially varying parameters},\ }\href
  {https://doi.org/https://doi.org/10.1016/j.physd.2005.01.022} {\bibfield
  {journal} {\bibinfo  {journal} {Physica D: Nonlinear Phenomena}\ }\textbf
  {\bibinfo {volume} {202}},\ \bibinfo {pages} {95} (\bibinfo {year}
  {2005})}\BibitemShut {NoStop}%
\bibitem [{\citenamefont {Ghosh}\ and\ \citenamefont {Ray}(2013)}]{SGDSR}%
  \BibitemOpen
  \bibfield  {author} {\bibinfo {author} {\bibfnamefont {S.}~\bibnamefont
  {Ghosh}}\ and\ \bibinfo {author} {\bibfnamefont {D.~S.}\ \bibnamefont
  {Ray}},\ }\bibfield  {title} {\bibinfo {title} {Chemical oscillator as a
  generalized rayleigh oscillator},\ }\href {https://doi.org/10.1063/1.4826169}
  {\bibfield  {journal} {\bibinfo  {journal} {The Journal of Chemical Physics}\
  }\textbf {\bibinfo {volume} {139}},\ \bibinfo {pages} {164112} (\bibinfo
  {year} {2013})},\ \Eprint
  {https://arxiv.org/abs/https://doi.org/10.1063/1.4826169}
  {https://doi.org/10.1063/1.4826169} \BibitemShut {NoStop}%
\bibitem [{\citenamefont {Mackey}\ and\ \citenamefont
  {Glass}(1977)}]{dynamicdise1}%
  \BibitemOpen
  \bibfield  {author} {\bibinfo {author} {\bibfnamefont {M.~C.}\ \bibnamefont
  {Mackey}}\ and\ \bibinfo {author} {\bibfnamefont {L.}~\bibnamefont {Glass}},\
  }\bibfield  {title} {\bibinfo {title} {Oscillation and chaos in physiological
  control systems},\ }\href {https://doi.org/10.1126/science.267326} {\bibfield
   {journal} {\bibinfo  {journal} {Science}\ }\textbf {\bibinfo {volume}
  {197}},\ \bibinfo {pages} {287} (\bibinfo {year} {1977})}\BibitemShut
  {NoStop}%
\bibitem [{\citenamefont {B{\'e}lair}\ \emph {et~al.}(1995)\citenamefont
  {B{\'e}lair}, \citenamefont {Glass}, \citenamefont {an~der Heiden},\ and\
  \citenamefont {Milton}}]{dynamicdise2}%
  \BibitemOpen
  \bibfield  {author} {\bibinfo {author} {\bibfnamefont {J.}~\bibnamefont
  {B{\'e}lair}}, \bibinfo {author} {\bibfnamefont {L.}~\bibnamefont {Glass}},
  \bibinfo {author} {\bibfnamefont {U.}~\bibnamefont {an~der Heiden}},\ and\
  \bibinfo {author} {\bibfnamefont {J.}~\bibnamefont {Milton}},\ }\bibfield
  {title} {\bibinfo {title} {Dynamical disease: identification, temporal
  aspects and treatment strategies of human illness},\ }\href@noop {}
  {\bibfield  {journal} {\bibinfo  {journal} {Chaos: An Interdisciplinary
  Journal of Nonlinear Science}\ }\textbf {\bibinfo {volume} {5}},\ \bibinfo
  {pages} {1} (\bibinfo {year} {1995})}\BibitemShut {NoStop}%
\bibitem [{\citenamefont {Da~Silva}\ \emph {et~al.}(2003)\citenamefont
  {Da~Silva}, \citenamefont {Blanes}, \citenamefont {Kalitzin}, \citenamefont
  {Parra}, \citenamefont {Suffczynski},\ and\ \citenamefont
  {Velis}}]{dynamicdise3}%
  \BibitemOpen
  \bibfield  {author} {\bibinfo {author} {\bibfnamefont {F.~L.}\ \bibnamefont
  {Da~Silva}}, \bibinfo {author} {\bibfnamefont {W.}~\bibnamefont {Blanes}},
  \bibinfo {author} {\bibfnamefont {S.~N.}\ \bibnamefont {Kalitzin}}, \bibinfo
  {author} {\bibfnamefont {J.}~\bibnamefont {Parra}}, \bibinfo {author}
  {\bibfnamefont {P.}~\bibnamefont {Suffczynski}},\ and\ \bibinfo {author}
  {\bibfnamefont {D.~N.}\ \bibnamefont {Velis}},\ }\bibfield  {title} {\bibinfo
  {title} {Epilepsies as dynamical diseases of brain systems: Basic models of
  the transition between normal and epileptic activity},\ }\href@noop {}
  {\bibfield  {journal} {\bibinfo  {journal} {Epilepsia}\ }\textbf {\bibinfo
  {volume} {44}},\ \bibinfo {pages} {72} (\bibinfo {year} {2003})}\BibitemShut
  {NoStop}%
\end{thebibliography}%


\begin{thebibliography}{10}

\bibitem{selkov}
E.~E. SEL'KOV, \emph{Eur. J. Biochem.} \textbf{1968}, \emph{4}, 79.

\bibitem{Goldbeter3255}
A.~Goldbeter, \emph{Proc. Natl. Acad. Sci. U. S. A.} \textbf{1973}, \emph{70},
  3255.

\bibitem{HYNNE2001121}
F.~Hynne, S.~Danø, P.~Sørensen, \emph{Biophys. Chem.} \textbf{2001},
  \emph{94}, 121.

\bibitem{Madsen}
M.~F. Madsen, S.~Danø, P.~G. Sørensen, \emph{FEBS J.} \textbf{2005},
  \emph{272}, 2648.

\bibitem{wolf2000effect}
J.~Wolf, R.~Heinrich, \emph{Biochem. J.} \textbf{2000}, \emph{345}, 321.

\bibitem{ZHANG2007112}
L.~Zhang, Q.~Gao, Q.~Wang, X.~Zhang, \emph{Biophys. Chem.} \textbf{2007},
  \emph{125}, 112.

\bibitem{mair2002spatio}
T.~Mair, C.~Warnke, S.~C. M{\"u}ller, \emph{Faraday Discuss.} \textbf{2001},
  \emph{120}, 249.

\bibitem{BAGYAN200567}
S.~Bagyan, T.~Mair, E.~Dulos, J.~Boissonade, P.~{De Kepper}, S.~C. Müller,
  \emph{Biophys. Chem.} \textbf{2005}, \emph{116}, 67.

\bibitem{LAVROVA2009127}
A.~Lavrova, S.~Bagyan, T.~Mair, M.~Hauser, L.~Schimansky-Geier,
  \emph{Biosystems} \textbf{2009}, \emph{97}, 127 .

\bibitem{Lavrova2009PhaseInflux}
A.~I. Lavrova, L.~Schimansky-Geier, E.~B. Postnikov, \emph{Phys. Rev. E}
  \textbf{2009}, \emph{79}, 1.

\bibitem{Verveykochaos}
D.~V. Verveyko, A.~Y. Verisokin, E.~B. Postnikov, \emph{Chaos} \textbf{2017},
  \emph{27}, 083104.

\bibitem{Murray2003MathematicalApplications}
J.~D. Murray, volume~2, Springer \textbf{2003}.

\bibitem{epstein1998introduction}
I.~R. Epstein, J.~A. Pojman, Oxford University Press \textbf{1998}.

\bibitem{reversibleselkoov}
P.~H. Richter, P.~Rehmus, J.~Ross, \emph{Prog. Theor. Phys.} \textbf{1981},
  \emph{66}, 385.

\bibitem{Rao2016NonequilibriumThermodynamics}
R.~Rao, M.~Esposito, \emph{Phys. Rev. X} \textbf{2016}, \emph{6}, 041064.

\bibitem{pkgg2}
P.~Kumar, G.~Gangopadhyay, \emph{Phys. Rev. E} \textbf{2021}, \emph{104},
  014221.

\bibitem{ertl1991oscillatory}
G.~Ertl, \emph{Science} \textbf{1991}, \emph{254}, 1750.

\bibitem{Reversalhetro}
T.-C. Li, B.-W.~a. Li, \emph{Chaos} \textbf{2013}, \emph{23}, 033130.

\bibitem{gilbert2010developmental}
S.~F. Gilbert, Sinauer Associates, Inc. \textbf{2010}.

\bibitem{PAGE200595}
K.~M. Page, P.~K. Maini, N.~A. Monk, \emph{Physica D} \textbf{2005},
  \emph{202}, 95.

\bibitem{Falasco2018InformationPatterns}
G.~Falasco, R.~Rao, M.~Esposito, \emph{Phys. Rev. Lett.} \textbf{2018},
  \emph{121}, 108301.

\bibitem{thermodynamicschemifcalwaves}
F.~Avanzini, G.~Falasco, M.~Esposito, \emph{J. Chem. Phys.} \textbf{2019},
  \emph{151}, 234103.

\bibitem{pkgg}
P.~Kumar, G.~Gangopadhyay, \emph{Phys. Rev. E} \textbf{2020}, \emph{101},
  042204.

\bibitem{pkgg3}
P.~Kumar, G.~Gangopadhyay, \emph{Phys. Rev. E} \textbf{2022}, \emph{105},
  034208.

\bibitem{experimentquasiperiodic}
F.~Argoul, A.~Arneodo, P.~Richetti, J.~C. Roux, \emph{J. Chem. Phys.}
  \textbf{1987}, \emph{86}, 3325.

\bibitem{Cross2009PatternSystems}
M.~Cross, H.~Greenside, Cambridge University Press \textbf{2009}.

\bibitem{strogatznonlinear}
S.~H. Strogatz, CRC press \textbf{2018}.

\bibitem{aranson2002world}
I.~S. Aranson, L.~Kramer, \emph{Rev. Mod. Phys.} \textbf{2002}, \emph{74}, 99.

\bibitem{Nicolis1995IntroductionScience}
G.~Nicolis, Cambridge University Press \textbf{1995}.

\bibitem{krylov1949introduction}
N.~M. Krylov, N.~N. Bogoliubov, Princeton University Press \textbf{1949}.

\bibitem{benjamin1967instability}
T.~B. Benjamin, \emph{Proc. R. Soc. A} \textbf{1967}, \emph{299}, 59.

\bibitem{lamomega1}
N.~Kopell, L.~N. Howard, \emph{Stud. Appl. Math.} \textbf{1973}, \emph{52},
  291.

\bibitem{LSODA}
L.~Petzold, \emph{SIAM J. Sci. Stat. Comput.} \textbf{1983}, \emph{4},
  136–148.

\bibitem{beatphoscillator}
H.~S. Lawson, G.~Holló, R.~Horvath, H.~Kitahata, I.~Lagzi, \emph{J. Phys.
  Chem. Lett.} \textbf{2020}, \emph{11}, 3014.

\bibitem{hirsch2012differential}
M.~W. Hirsch, S.~Smale, R.~L. Devaney, Academic press \textbf{2012}.

\bibitem{postoperiodic}
D.~V. Verveyko, A.~Y. Verisokin, E.~B. Postnikov, \emph{Chaos} \textbf{2017},
  \emph{27}, 083104.

\bibitem{SGDSR}
S.~Ghosh, D.~S. Ray, \emph{J. Chem. Phys.} \textbf{2013}, \emph{139}, 164112.

\bibitem{dynamicdise1}
M.~C. Mackey, L.~Glass, \emph{Science} \textbf{1977}, \emph{197}, 287.

\bibitem{dynamicdise2}
J.~B{\'e}lair, L.~Glass, U.~an~der Heiden, J.~Milton, \emph{Chaos}
  \textbf{1995}, \emph{5}, 1.

\bibitem{dynamicdise3}
F.~L. Da~Silva, W.~Blanes, S.~N. Kalitzin, J.~Parra, P.~Suffczynski, D.~N.
  Velis, \emph{Epilepsia} \textbf{2003}, \emph{44}, 72.

\bibitem{SCHUTZE2010104}
J.~Schütze, J.~Wolf, \emph{Biosystems} \textbf{2010}, \emph{99}, 104.

\bibitem{PETTY2006217}
H.~R. Petty, \emph{Biosystems} \textbf{2006}, \emph{83}, 217.

\bibitem{PURVIS2013945}
J.~E. Purvis, G.~Lahav, \emph{Cell} \textbf{2013}, \emph{152}, 945.

\bibitem{BEHAR2010684}
M.~Behar, A.~Hoffmann, \emph{Curr. Opin. Genet. Dev.} \textbf{2010}, \emph{20},
  684.

\end{thebibliography}
	
\clearpage

%\section*{Entry for the Table of Contents}   
%
%\noindent\rule{11cm}{2pt}
%\begin{minipage}{11cm}
%	\includegraphics[width=11cm]{TOClatest.eps}
%\end{minipage}
%\begin{minipage}{11cm}
%	\large\textsf{Different temporal behaviors related to spreading the inhomogeneous control parameter over the spatial domain are illustrated. Traveling wave direction change occurs only with slight asymmetry in the control parameter profile.\\}
%\end{minipage}
%\noindent\rule{11cm}{2pt}
%
%\vspace{13cm}
%
%
%Premashis Kumar's personal Twitter handle: @premashis$\_$kumar	
\end{document}